\documentstyle[12pt,a4wide,epsfig]{article}
\begin{document}
\renewcommand{\topfraction}{1}
\renewcommand{\bottomfraction}{1}
\renewcommand{\textfraction}{0}
\hfill \parbox[t]{3cm}{WUB 98-27 \\ \hfill HLRZ1998-54}
\begin{center}
{\Large\bf The Pion-Nucleon  $\sigma$-Term \\
 with Dynamical Wilson Fermions}
\vskip 1cm
SESAM Collaboration 
\vskip 0.5cm
S.~G\"usken$^a$, P.~Ueberholz$^a$, J.~Viehoff$^a$ \\

N.~Eicker$^b$, P.~Lacock$^b$, T.~Lippert$^a$, K.~Schilling$^{a,b}$, 
A.~Spitz$^b$, T.~Struckmann$^a$
\vskip 5mm
\normalsize\it{$^a$ Bergische Universit\"at Wuppertal, Fachbereich
  Physik,
  42097 Wuppertal, Germany} \\
\normalsize\it{$^{b}$ HLRZ Forschungszentrum J\"ulich, and DESY,
  Hamburg, 52425 J\"ulich, Germany} 
\abstract{\small We calculate
  connected and disconnected contributions to the flavour singlet
  scalar density amplitude of the nucleon in a full QCD lattice
  simulation with $n_f=2$ dynamical Wilson fermions at $\beta=5.6$ on
  a $16^3 \times 32$ lattice. We find that both contributions are of
  similar size at the light quark mass. We
  arrive at the estimate $\sigma_{\pi N} = 18(5)$MeV.  Its smallness
  is directly related to the apparent decrease of $u$, $d$ quark
  masses when unquenching QCD lattice simulations.  The $y$ parameter
  can be estimated from a semi-quenched analysis, in which there are
  no strange quarks in the sea, the result being $y=0.59(13)$.  }
\end{center}
 
\section{Introduction} 
\begin{figure}
\begin{center}
\epsfxsize=12cm
\epsfbox{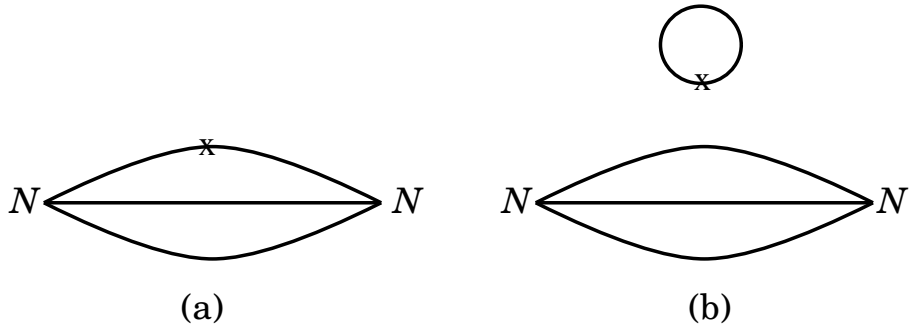}
\end{center}   
\caption{\label{fig_schema_con_dis}{\it Connected (a) and disconnected (b)
contributions to the scalar density amplitude of a nucleon. Please note that
all quark lines, including the quark loop, are connected by infinitely
many gluon lines and virtual quark loops. 
 }} 
\end{figure}

\normalsize  
The pion-nucleon $\sigma$-term is defined as  the flavour singlet 
scalar density amplitude of the nucleon, multiplied by the light quark mass
$m_{ud}$
\begin{equation}
\sigma_{\pi N} = m_{ud} \langle N|\bar{u}u + \bar{d}d|N \rangle
\quad , \quad m_{ud} = 1/2(m_u + m_d)
\;.
\label{eq_sigma_def}
\end{equation}
Our motivation to study this quantity in a full QCD simulation is
twofold. First of all $\sigma_{\pi N}$ provides a direct measure
of the explicit chiral symmetry breaking in QCD. A comparison
of its experimental value with a (first principles) QCD calculation  
is therefore of great importance for the understanding of the
chiral properties of the strong interaction. Secondly,
the scalar density amplitude receives contributions from both
connected and disconnected (vacuum polarization) diagrams. This is
shown in fig.\ref{fig_schema_con_dis}.
In a flavour singlet combination, such as $\sigma_{\pi N}$, the latter
are simply added.
Therefore, one expects sizeable contributions from those processes, which
most likely will depend on the details of the vacuum structure.  
Given this, the pion-nucleon $\sigma$-term provides an ideal opportunity
to disclose the impact of sea quarks on nucleon properties.

On the other hand the determination of the experimental value of
$\sigma_{\pi N}$ is by no means straightforward as it requires quite a
bit of theoretical input. For instance it implies the use of
pion-nucleon scattering data at the unphysical Cheng-Dashen point.  A
careful analysis of the extrapolation procedures has been performed by
Gasser, Leutwyler and Sainio\cite{gasser_nsigma} by means of
dispersion relations and chiral perturbation theory. They found
$\sigma_{\pi N} \simeq 45 \mbox{MeV}$.  A consistent value,
{$\sigma_{\pi N} = 48\pm 10$MeV}, has been obtained more recently in
the framework of heavy baryon chiral perturbation theory by the
authors of ref.\cite{meissner_nsigma}.
   
Additional information can be drawn  from the baryon octet mass
splittings. The flavour octet quantity
\begin{equation}
\sigma_0 = 
\frac{m_u+m_d}{2} \langle N|\bar{u}u + \bar{d}d - 2\bar{s}s|N \rangle
\end{equation}
is related to $\sigma_{\pi N}$ by
\begin{equation}
\sigma_{\pi N} = \frac{\sigma_0}{1 - y}\;,\;
y=
\frac{2\langle N|\bar{s}s |N \rangle}{\langle N|\bar{u}u+\bar{d}d |N
  \rangle} \;.
\label{eq_y_def}
\end{equation}
The analysis of the baryon octet mass splittings in first order chiral
perturbation theory yields $\sigma_0 \simeq 25$MeV. It has been
pointed out however\cite{gasser_2nd_order}, that corrections due to
terms $\propto m_s - 1/2(m_u+m_d)^2$ may enhance this value on a
$(20-30)\%$ level. Relating these findings with the results of the
$\pi N$ scattering analysis leads to the estimate $y \simeq 0.2 -
0.4$.  This implies that strange quark loops decrease the value of
$\sigma_0$ by $20 - 40 \%$. Naively, i.e. under the assumption of an
approximate $SU(3)$ flavour symmetry of the disconnected parts of the
nucleon scalar density, one would expect light quark loops to
contribute similarly to the pion-nucleon $\sigma$-term.

$\sigma_{\pi N}$ as well as $y$ have been studied recently in quenched
lattice simulations by Fukugita et al.\cite{japan_nsigma} 
and Dong et al.\cite{liu_nsigma}. 
Both calculations find a value for $\sigma_{\pi N}$ consistent with
the `experimental' results quoted above, and a rather large
ratio \newline
$<N|\bar{u}u+\bar{d}d|N>_{disc}/<N|\bar{u}u+\bar{d}d|N>_{con} \simeq
2/1$.
     
Furthermore, ref.\cite{japan_nsigma} estimates $y = 0.66(15)$.  Given
the value of $2:1$ for the ratio of disconnected to connected
contributions, this is exactly the value one would expect from the
assumption of flavour symmetry of the disconnected contributions.
Such a finding appears plausible since the quenched QCD vacuum is 
sensitive neither to quark flavour nor  mass.

The authors of ref.\cite{liu_nsigma} on the other hand find a much
smaller value, $y=0.36(3)$. But this is mostly due to their use of
phenomenologically inspired ans\"atze for the extrapolation in the
quark mass and for the renormalization of disconnected contributions.
We will come back to this point later.  Moreover, their data sample
being very limited, one might have  doubts on the reliability of
their error analysis.

Apart from this, it is by no means obvious that quenched lattice
simulations are at all suited  to yield sound first principle QCD
estimates for $\sigma_{\pi N}$ and $y$. 
For the quenched approximation neglects internal quark loops in
the vacuum field configurations. As sea quark loops are essential
in the calculation of disconnected contributions, the quenched
approximation appears to be inconsistent and might introduce
a serious systematic bias to $\sigma_{\pi N}$ and $y$.   

It is therefore of utmost importance to study these quantities in full
QCD lattice simulations. Apart from the issue  of the numerical
value of $\sigma_{\pi N}$ one might learn about the physics  of sea
quarks in QCD from the relative weight of disconnected to connected
contributions and the amount of flavour symmetry breaking in the
disconnected sector.

We emphasize that the size of the $y$ parameter is determined  by
these quantities: a large ratio of disconnected to connected parts and a
high degree  of symmetry breaking impose  a low value on  $y$.
      
The calculation of disconnected correlations functions in lattice QCD
requires a high statistics of gauge field configurations, since such
correlators receive contributions only from vacuum fluctuations.  Even
with a `state of the art' statistics of $O(200)$ configurations one
needs elaborate analysis techniques to enhance the signal to noise
ratio.  Previous exploratory full QCD simulations
\cite{mtc_full_sigma,gupta_full_sigma} have therefore not been
sensitive enough to resolve a possible unquenching effect.

In this paper we present the results of a full QCD lattice simulation
with $n_f=2$ quark flavours of (mass degenerate) Wilson fermions at
$\beta=5.6$ and a lattice volume of $n_s^3\times n_t =16^3\times 32$
points. This corresponds to a lattice cutoff $a_{\rho}^{-1} \simeq
2.3$GeV and a spatial length $L \simeq 1.4$fm.  We have generated 200
statistically independent vacuum configurations at each of our 4
values of the sea quark mass, which correspond to
$m_{\pi}/m_{\rho}=0.833(3),0.809(15),0.758(11)$ and 0.686(11). Details
of the simulation as well as the analysis of the light hadron spectrum
can be found in \cite{sesam_light_spectrum}.

The paper is organized as follows. In chapter 2 we will define
the lattice correlators and explain our 
methods to extract the connected
and the disconnected scalar density amplitudes of the nucleon.
In chapter 3 the raw results are presented and the quality of our
signals as they emerge from different analysis methods is discussed.
The extrapolation to light quarks is performed in chapter 4. Here we 
also obtain our results for $\sigma_{\pi N}$ and $y$.
Finally, discussion and conclusions are given in chapter 5.     

\section{Analysis Setup}
   
\subsection{Ratio Methods}

In order to optimize the signal to noise ratio 
and to study the systematics of different procedures
we have applied several analysis methods 
both to the connected and to the disconnected parts
of the scalar density matrix element.

\subsubsection{Global Summation Method}
The  standard procedure is to consider the ratio\cite{maiani_ratio}
\begin{eqnarray}
\lefteqn{R^{SUM}(t) =} \label{eq_sum_meth_def} \\
 && \frac{\sum_{\vec{x}}
\langle N^{\dagger}(\vec{0},0) \sum_{\vec{y},y_0}\left[\bar{q}q\right]
(\vec{y},y_0)
N(\vec{x},t) \rangle }
{\sum_{\vec{x}}\langle N^{\dagger}(\vec{0},0) N(\vec{x},t) \rangle}
-  \langle\sum_{\vec{y},y_0}\left[\bar{q}q\right](\vec{y},y_0)
 \rangle\;. \nonumber
\end{eqnarray} 
$N$  denotes an interpolating operator for the nucleon.
With the help of the generating functional formalism of the path
integral, and using the Feynman-Hellmann theorem, one obtains 
\begin{equation}
R^{SUM}(t) = A + \langle N|\bar{q}q|N \rangle\,t \;, 
\label{eq_sum_meth_asympt}
\end{equation}   
provided the nucleon is in its ground state. Naively one would expect
this requirement to hold at large time distances $t$. However, due
to the summation over all time positions of $\bar{q}q$ in 
eq.\ref{eq_sum_meth_def}, $R^{SUM}$ might even then be contaminated by
nucleon excitations. Furthermore, the ratio receives contributions
from quark loops at distances $y_0 >> t$. Those do not add
substantially to the signal, but, especially for the disconnected
part, might enhance the noise.  It is therefore advantageous to undo
the summation over $y_0$ by evaluating the signals at definite values
$y_0$, in the range $0 \ll y_0 \ll t$.
 
\subsubsection{Plateau Density Method}

The local density ratio 
\begin{eqnarray}
\lefteqn{R^{PLA}(t,y_0) =} \label{eq_plateau_meth_def} \\
 && \frac{\sum_{\vec{x}}
\langle N^{\dagger}(\vec{0},0) \sum_{\vec{y}}\left[\bar{q}q\right]
(\vec{y},y_0)
N(\vec{x},t) \rangle }
{\sum_{\vec{x}}\langle N^{\dagger}(\vec{0},0) N(\vec{x},t) \rangle}
- \langle\sum_{\vec{y}}\left[\bar{q}q\right](\vec{y},y_0)
 \rangle \nonumber
\end{eqnarray} 
allows for an isolation of the proton ground state with respect to both
$t$ and $y_0$. 
Its asymptotic time dependence 
can be evaluated in the transfer matrix formalism, without recourse to
the Feynman-Hellmann theorem. 
For 
$0 \ll y_0 \ll t$ the ratio becomes independent of $t$ and $y_0$, and 
one obtains\footnote{This form is valid for an
infinitely  extended lattice in time.} 
\begin{equation}
R^{PLA}(t,y_0) = \langle N |\bar{q}q|N \rangle \;.
\label{eq_plateau_meth_asympt}
\end{equation}
Thus, the ground state signature of $R^{PLA}$ is a plateau instead of a
linear rise. The height of the plateau will tell us about the scalar
density nucleon matrix element.

We will demonstrate below  that this method works well for the
connected parts. However, one looses the advantage to gain statistics
by summing over $y_0$, which becomes a crucial issue for the disconnected
part.  For the calculation of the latter we have therefore decided to
use a slightly modified technique  which still accumulates data with
respect to $y_0$, but in a region characterized by ground state
dominance of the signals.

\subsubsection{Plateau Accumulation Method}

The plateau accumulation method (PAM) combines
the advantages of the 
global summation and of the plateau density techniques. It is defined by
\begin{equation}
R^{PAM}(t,\Delta t_0,\Delta t)
 = \sum_{y_0=\Delta t_0}^{t-\Delta t} R^{PLA}(t,y_0)\;,
\label{eq_mplateau_meth_def}
\end{equation}  
with $1 \leq \Delta t,\Delta t_0 \leq t$. 
The asymptotic time dependence is given by
\begin{equation}
R^{PAM}(t,\Delta t_0,\Delta t) =
B + \langle N |\bar{q}q|N \rangle (t-\Delta t - \Delta t_0) \;.
\label{eq_mplateau_meth_asympt}
\end{equation}

\subsection{Numerical Evaluation}

\subsubsection{Connected Contributions}

We compute the numerator of eq.\ref{eq_sum_meth_def} (summation
method) with the standard insertion technique \cite{maiani_ratio}.
This method is advantageous if one has to sum over all time positions
$y_0$. The additional  effort, on top of the standard  quark propagator
calculation, is just to compute a modified quark propagator 
$\tilde{\Delta}(x,0)$, defined as the solution to    
\begin{equation}
M \tilde{\Delta}(x,0) = \Delta(x,0) \;.
\label{eq_insert_standard}
\end{equation}
Here $M$ is the fermion matrix and $\Delta$ the quark
propagator. Graphically, the modified quark propagator corresponds
to the quark line with a cross in  fig. \ref{fig_schema_con_dis}a.

The standard insertion technique can be in principle applied also to
the numerator of eq.\ref{eq_plateau_meth_def} (plateau method).
However, as one avoids the summation over $y_0$ in this case, one
would have to solve eq.\ref{eq_insert_standard} $n_t$ times.  Instead,
we use the  advanced insertion technique proposed by Martinelli and
Sachrajda\cite{new_ins}. Here, one keeps  the nucleon sink and source
at the largest possible time separation\footnote{For (anti)periodic
  boundary conditions, the largest possible time distance is
  $n_t/2$.}, and computes  an advanced propagator by solving the
equation
\begin{equation}
\epsfxsize=11cm
\epsfbox{martinelli_insertion.eps}\nonumber
\label{eq_martinelli_insertion}
\end{equation}
Note that the r.h.s. is just a combination of standard quark
propagators. The 3-point correlation is then given by the product
of the advanced and the standard quark propagator. The
position in time, $y_0$, of the quark density $\bar{q}q$ is not fixed
here and can be varied without additional cost.  
 
\subsubsection{Disconnected Contributions}

The determination of the quark loop contributions to the 
scalar density matrix element requires the calculation of the
trace of the quark propagator
\begin{equation}
L(y_0) = 
\sum_{\vec{y},\alpha,a}[\bar{q}q](\vec{y},y_0,\alpha,a;\vec{y},y_0,\alpha,a)
= Tr \Delta(y_0;y_0) \;.
\end{equation}
$\alpha$ and $a$ denote Dirac and colour degrees of freedom.
An exact determination of $L$ with conventional Krylov subspace
methods would be prohibitively expensive since one would have 
to apply such an algorithm $n_s^3 \times n_t$ times.
Instead one has to rely on approximate methods like the volume source 
technique\cite{japan_volume_source} and
the stochastic estimator technique with Gaussian\cite{bitar_gauss} and with
$Z_2$\cite{liu_z2} noise. The latter allow to control
the accuracy of the estimator on
each gauge configuration. 

It has been demonstrated
recently\cite{sesam_disc} that the stochastic estimator technique
with (complex)  $Z_2$ noise is superior for our lattice setup. 
Therefore we used this method here, albeit with small modifications
which allow for the determination of
\begin{equation}
L(y_0,\beta',\beta) = 
\sum_{\vec{y},a} \Delta(\vec{y},y_0,a,\beta';\vec{y},y_0,a,\beta) \;,
\end{equation}
where the Dirac indices are not contracted.  Thus, the results can be
re-used to calculate vacuum loops other than scalars.  For
completeness we sketch the relevant formulae of this {\it spin
  explicit method} here.

On a given configuration and for each estimate we choose $N_E$ complex
$Z_2$ random vectors $\eta(\vec{y},y_0,a,\alpha)$, with
$n_s^3\times n_t \times 4 \times 3$ entries. Each component of $\eta$
has the properties
\begin{equation}
\eta^*(i)\eta(i) = 1 \quad , \quad \langle \eta^*(i)\eta(j) \rangle =
0\;\; \mbox{for}\; i\neq j \;.
\label{eq_z2_properties}
\end{equation}
The brackets of the right equation denote the average over (infinitely
many) stochastic estimates. From $\eta$ we compose 4 {\it spin explicit} 
random vectors
\begin{equation}
\eta^{\beta}(\vec{y},y_0,a,\alpha) = 
\eta(\vec{y},y_0,a,\alpha) \delta_{\alpha,\beta} \quad,
\beta=0,1,2,3\; .
\end{equation}  
Note that there is no sum over $\alpha$ on the r.h.s. .
The vector $\Delta\eta^{\beta}$ is then obtained as the solution to
\begin{equation}
M [\Delta \eta^{\beta}] = \eta^{\beta} \; .
\end{equation}
According to eq.\ref{eq_z2_properties}, the product 
\begin{eqnarray}
\lefteqn{P(t_0,\beta',\beta) = (\tilde{\eta}^{\beta',t_0})^\dagger \Delta \eta^{\beta}
\;, } \\
&& \mbox{with} \quad \tilde{\eta}^{\beta',t_0}(\vec{y},y_0,a,\alpha) = 
\eta^{\beta}(\vec{y},y_0,a,\beta)\delta_{y_0,t_0}\delta_{\alpha,\beta'}
\nonumber
\end{eqnarray}
converges in the limit $N_E \rightarrow \infty$  on each gauge
configuration to
\begin{equation}
   \langle P(t_0,\beta',\beta) \rangle =  L(y_0,\beta',\beta) \;. 
\end{equation}
In this work we have used 100 stochastic estimates per configuration.
We have checked that this suffices to reach the asymptotic
region.

\section{Raw Data}

\begin{figure}[htb]
\begin{center}
\vskip -3.0cm
\noindent\parbox{15.0cm}{
\parbox{7.0cm}{\epsfxsize=7.0cm\epsfbox{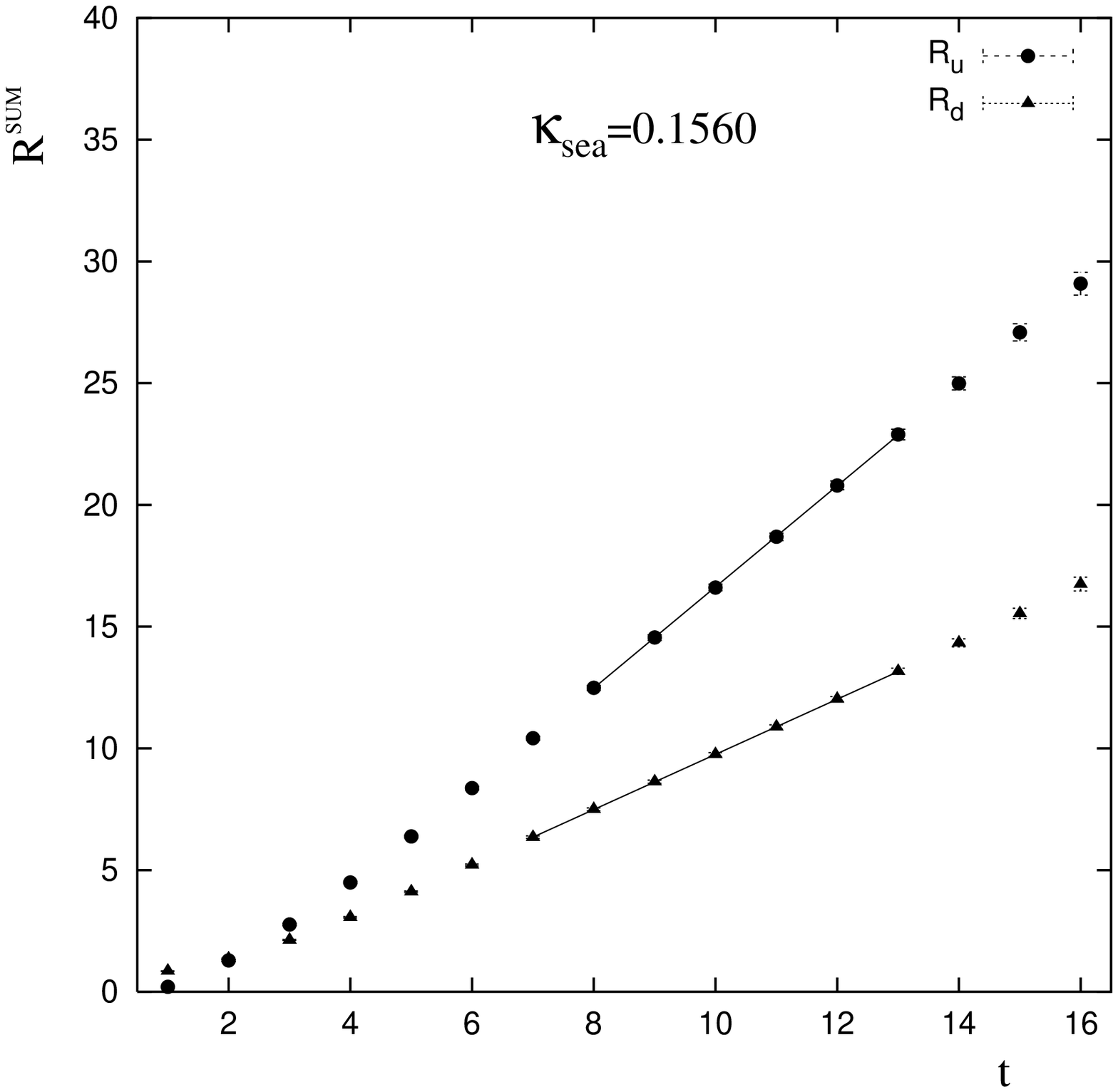}}
\parbox{7.0cm}{\epsfxsize=7.0cm\epsfbox{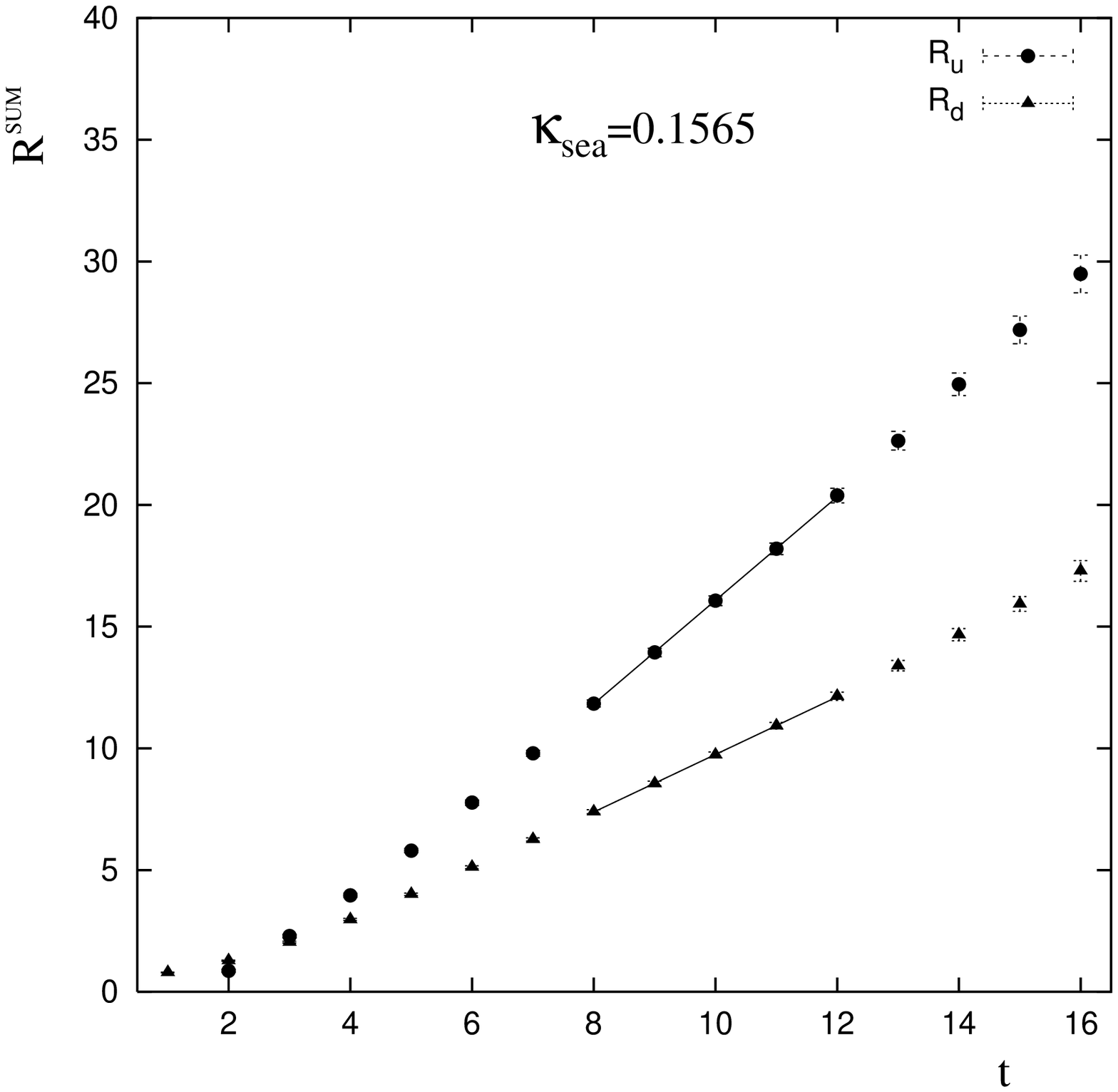}}
\\ \linebreak
\vskip -4.0cm
\parbox{7.0cm}{\epsfxsize=7.0cm\epsfbox{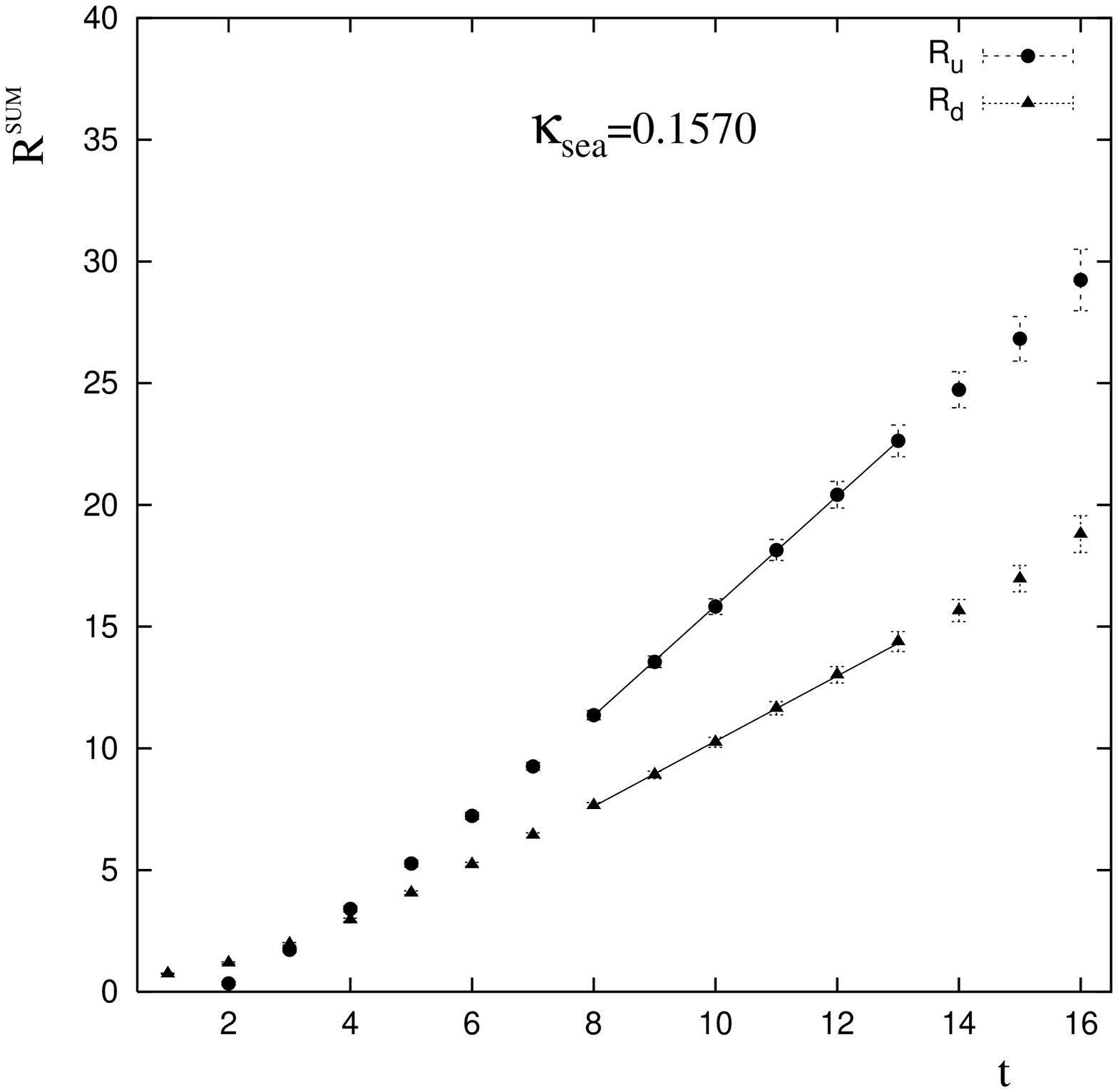}} 
\parbox{7.0cm}{\epsfxsize=7.0cm\epsfbox{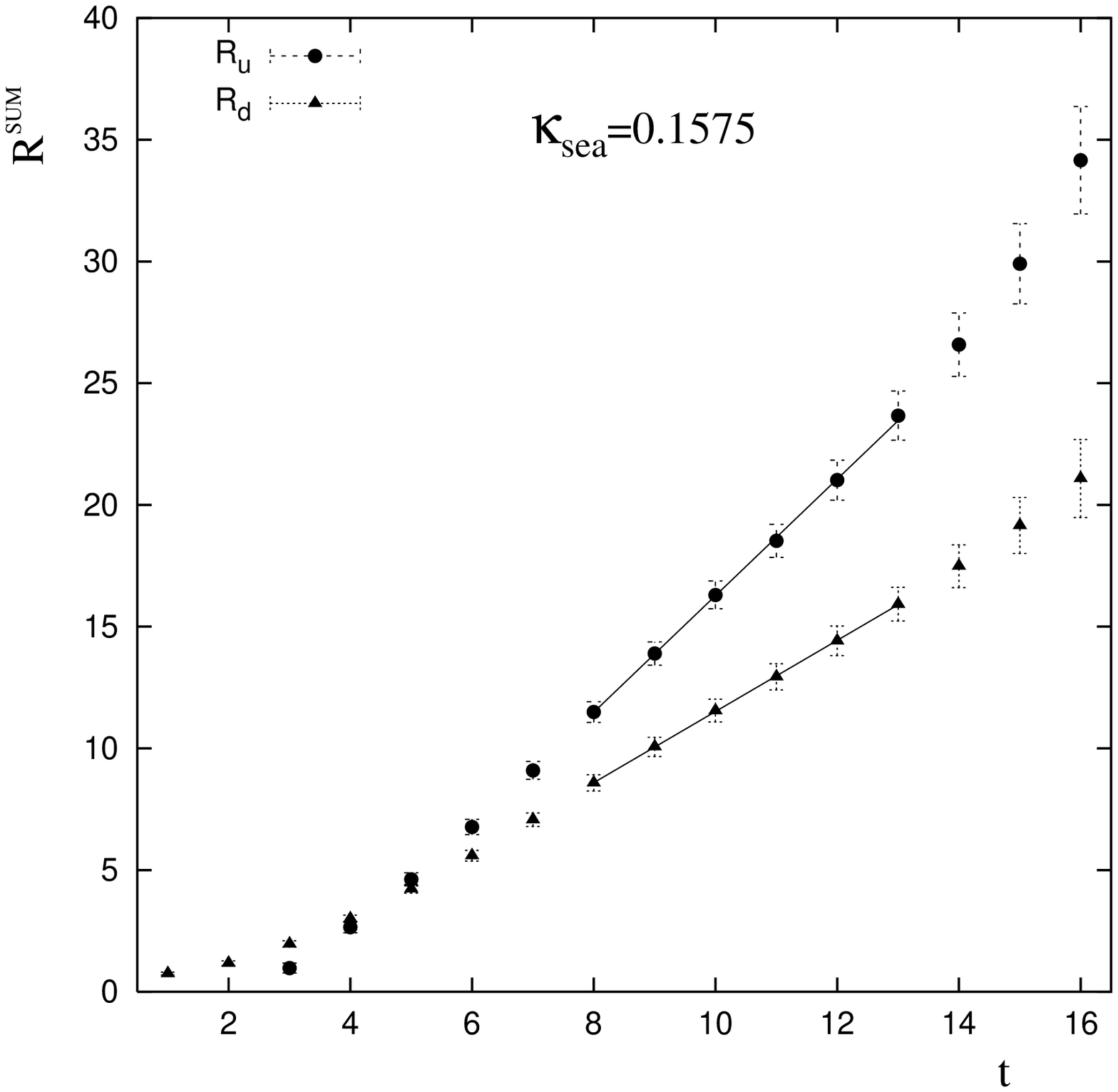}}\\
}
\caption{\label{fig_rawdata_rsum} {\it Summation method: The raw data $R_u$
  and $R_d$ for the  connected amplitudes $C_{u,d}$ at our 
sea quark masses. The fits (range and value) are indicated by solid lines.}}
\end{center}
\end{figure}

\begin{figure}[htb]
\begin{center}
\vskip -3.0cm
\noindent\parbox{15.0cm}{
\parbox{7.0cm}{\epsfxsize=7.0cm\epsfbox{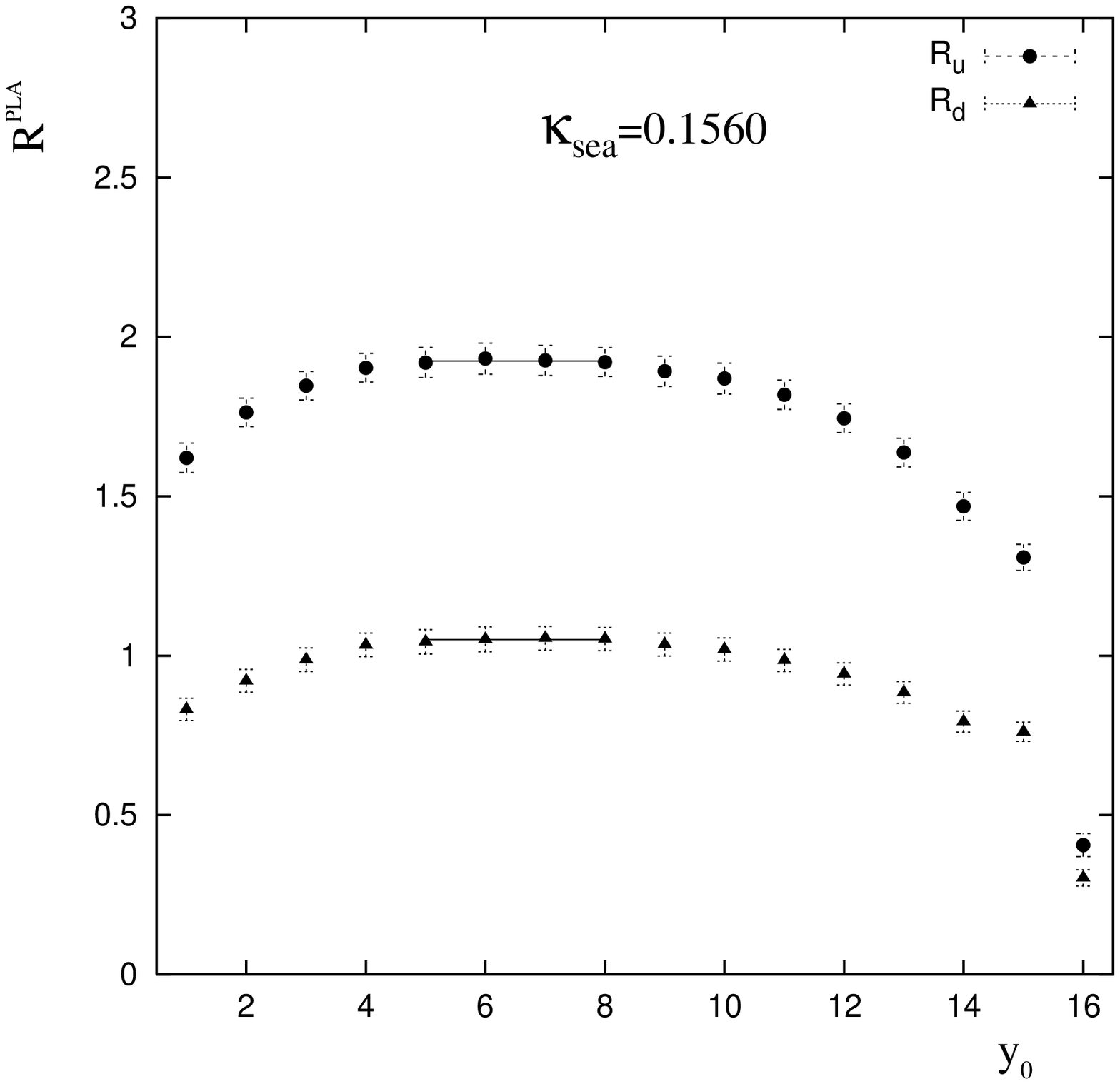}}
\parbox{7.0cm}{\epsfxsize=7.0cm\epsfbox{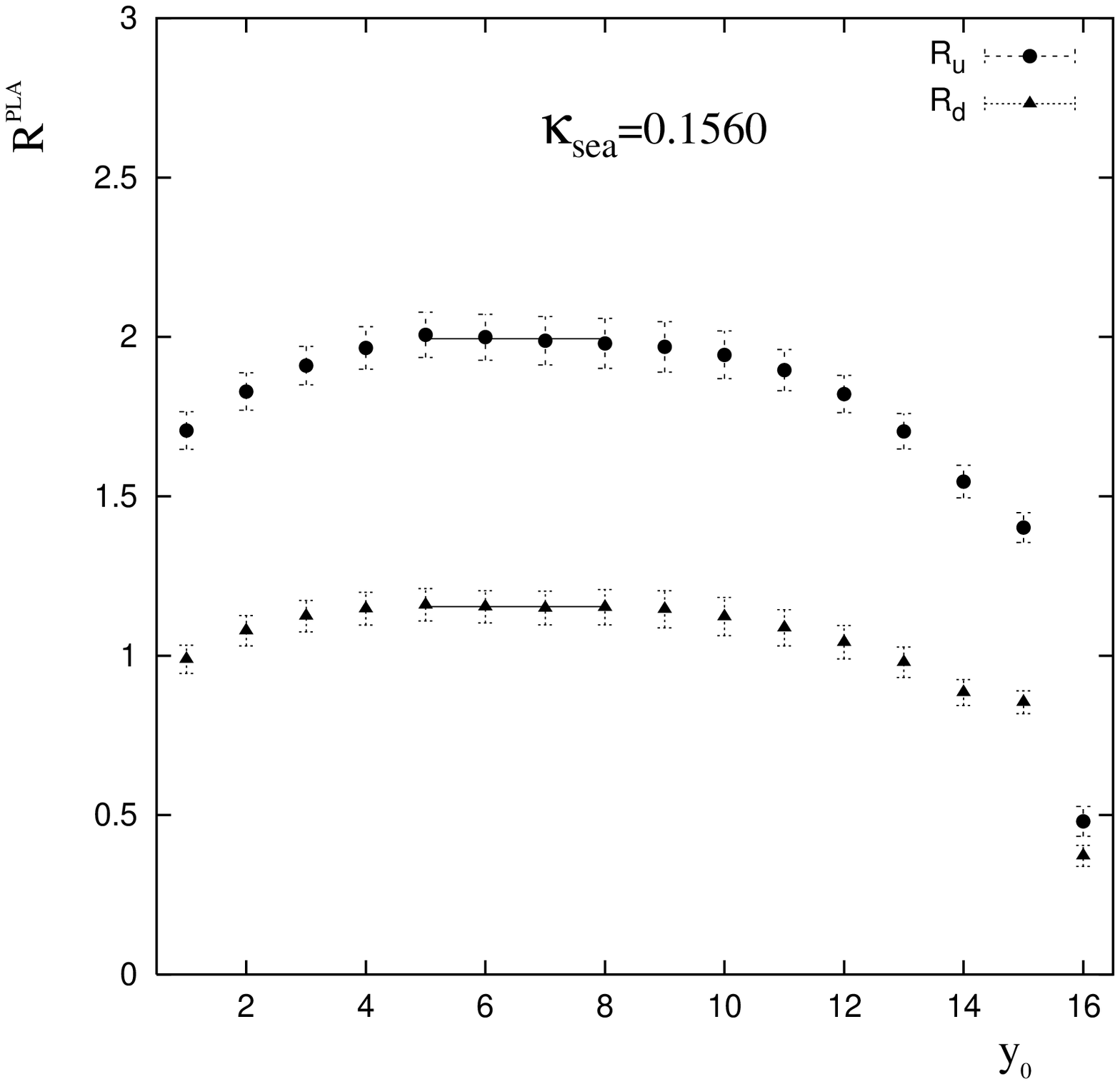}}
\\ \linebreak
\vskip -4.0cm
\parbox{7.0cm}{\epsfxsize=7.0cm\epsfbox{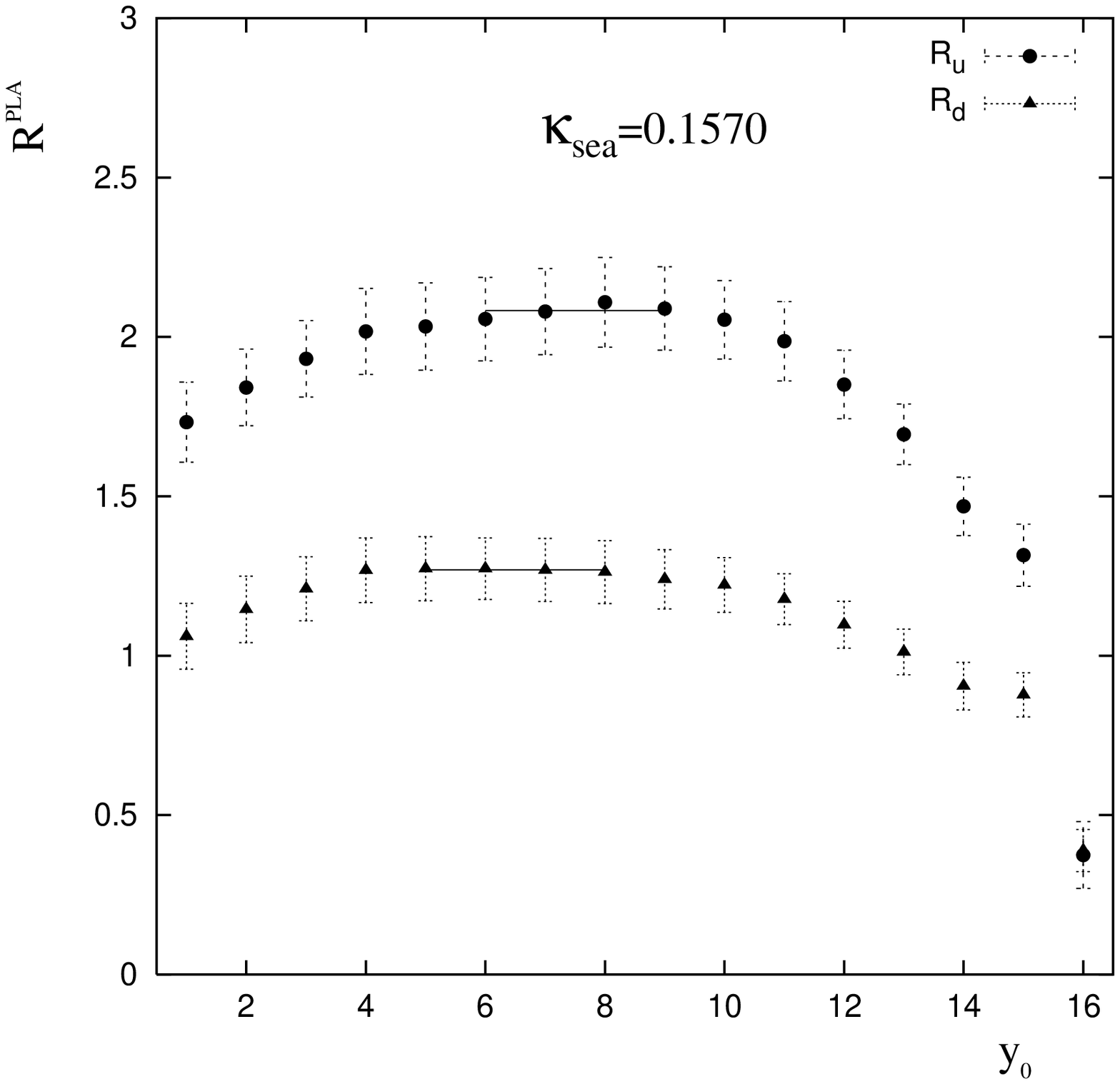}} 
\parbox{7.0cm}{\epsfxsize=7.0cm\epsfbox{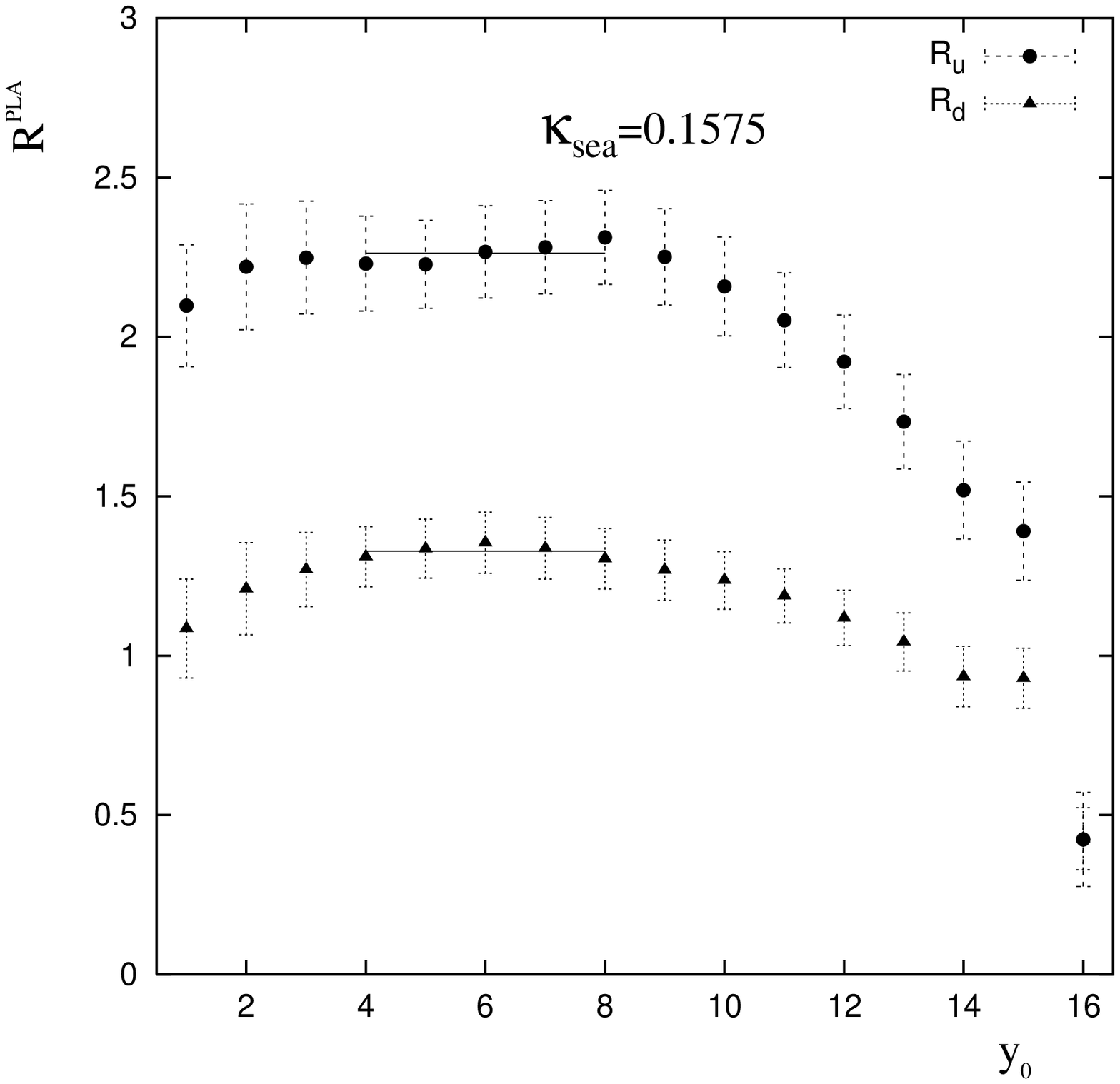}}\\
}
\caption{\label{fig_rawdata_rpla} {\it Plateau method: The raw data $R_u$
and $R_d$ for the  connected amplitudes $C_{u,d}$ at our
sea quark masses. The fits (range and value) are indicated by solid lines.}}
\end{center}
\end{figure}

Figs.\ref{fig_rawdata_rsum} and \ref{fig_rawdata_rpla}
display the ratios $R^{SUM}$
and $R^{PLA}$ of the connected contributions for our four quark masses.
On each graph we show the data for both,  the (scalar)
interactions of the $u$ and of the $d$ quarks of the proton.

 Note that $R^{PLA}$ is plotted in fig.\ref{fig_rawdata_rpla} as a
function of the (time) position of the scalar interaction $y_0$. 
The time separation  of proton sink and source is fixed at $t=16$.
  
All ratios exhibit clear signals, even for the smallest value of      
our quark masses. From fits according to eqs.\ref{eq_sum_meth_asympt}
and \ref{eq_plateau_meth_asympt} we extract the connected parts of
the scalar density matrix element of the proton. These are
listed in tab.\ref{tab_raw_connected}.
\begin{table}[ht]
\begin{center}
\begin{tabular}{|c|c|c|c|c|}
\hline 
           &         &          &          &           \\
Method     & $\kappa$ & $C_u$    & $C_d$    & $C_{u+d}$ \\
           &         &          &          &           \\ \hline
           & 0.1560  & 2.08(4)  & 1.13(2)  & 3.21(5)   \\
 SUM       & 0.1565  & 2.13(6)  & 1.18(4)  & 3.31(9)   \\
           & 0.1570  & 2.26(11) & 1.34(6)  & 3.60(17)  \\
           & 0.1575  & 2.40(16) & 1.46(13) & 3.86(28)  \\ \hline
           & 0.1560  & 1.92(5)  & 1.05(4)  & 2.98(8)   \\
 PLATEAU   & 0.1565  & 1.99(8)  & 1.15(5)  & 3.15(13)  \\
           & 0.1570  & 2.08(13) & 1.27(11) & 3.35(24)  \\
           & 0.1575  & 2.26(14) & 1.33(11) & 3.59(23)  \\ \hline
\end{tabular}
\caption{\label{tab_raw_connected}{\it Lattice results
(unrenormalised) of the connected amplitude 
$C_q = \langle P(\kappa_{sea})
|\bar{q}q(\kappa_{sea})|P(\kappa_{sea})\rangle_{con}$.}}
\end{center}
\end{table}

Both methods yield consistent results within statistical errors.
It appears however, that the summation method systematically leads
to slightly larger values than the plateau method. We attribute this to a 
small contamination of the SUM results with excited proton
contributions. We will therefore use the data from the plateau method
for our final analysis.
\begin{figure}[htb]
\begin{center}
\vskip -3.0cm
\noindent\parbox{15.0cm}{
\parbox{7.0cm}{\epsfxsize=7.0cm\epsfbox{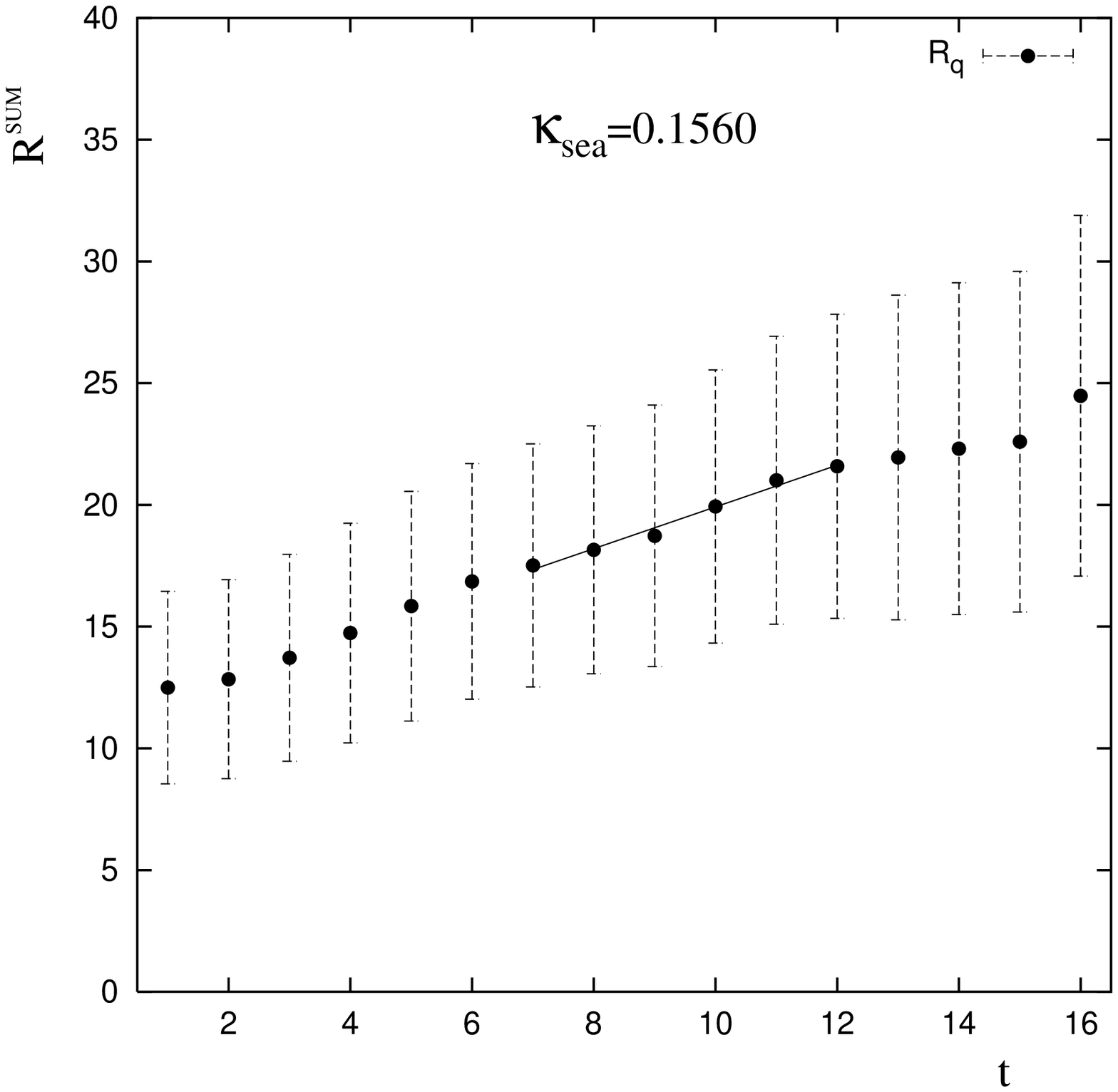}}
\parbox{7.0cm}{\epsfxsize=7.0cm\epsfbox{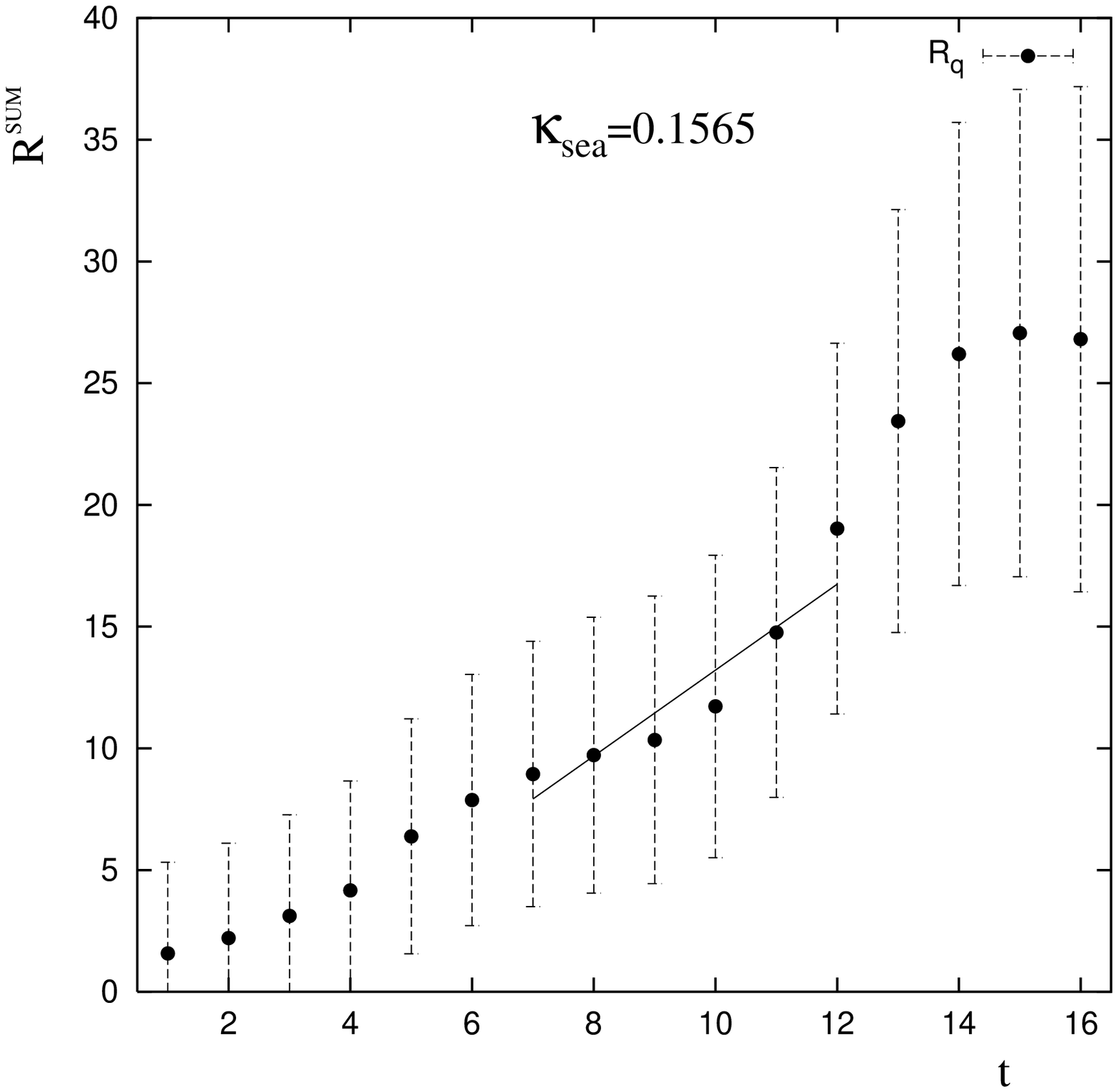}}
\\ \linebreak
\vskip -4.0cm
\parbox{7.0cm}{\epsfxsize=7.0cm\epsfbox{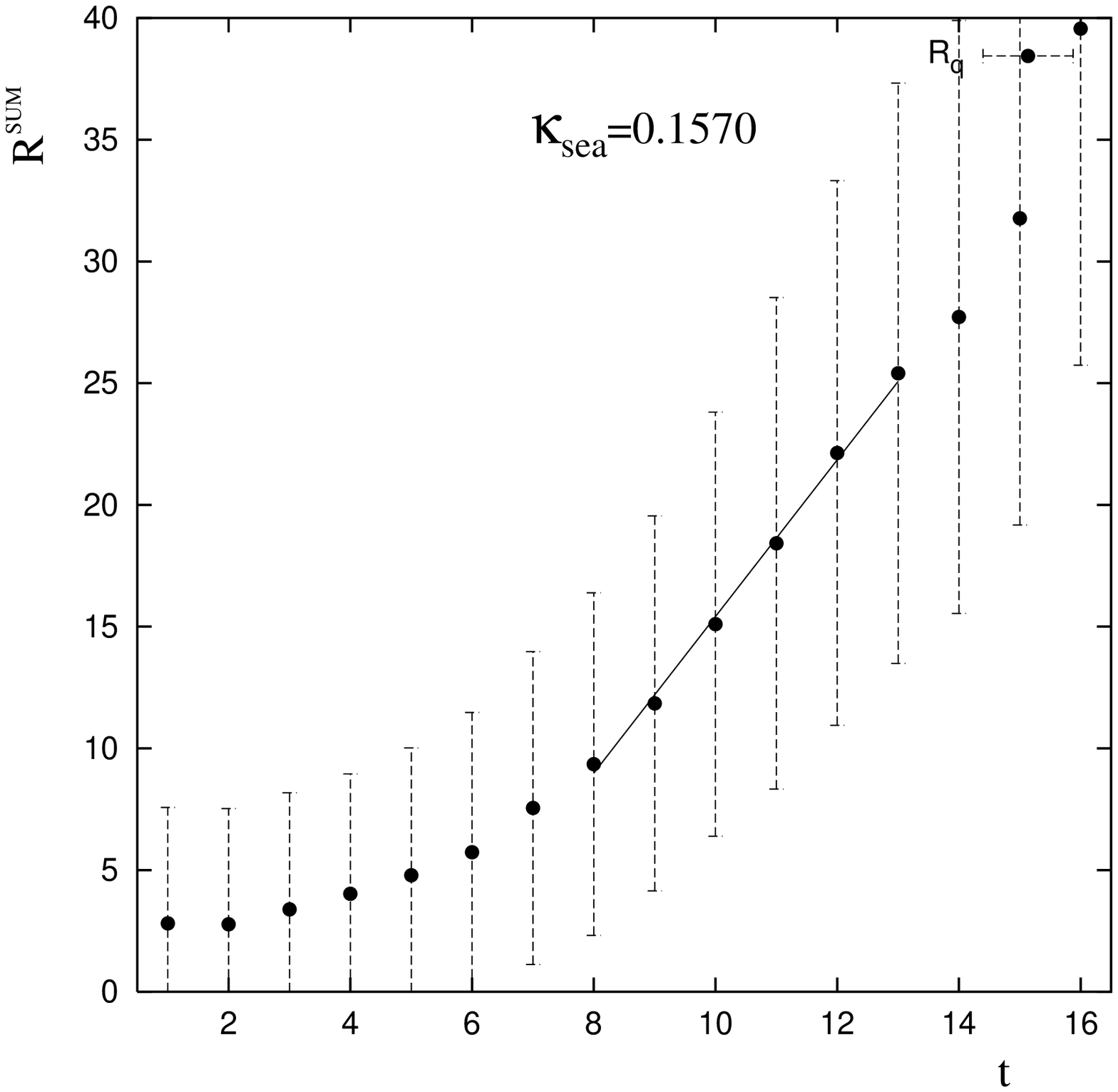}} 
\parbox{7.0cm}{\epsfxsize=7.0cm\epsfbox{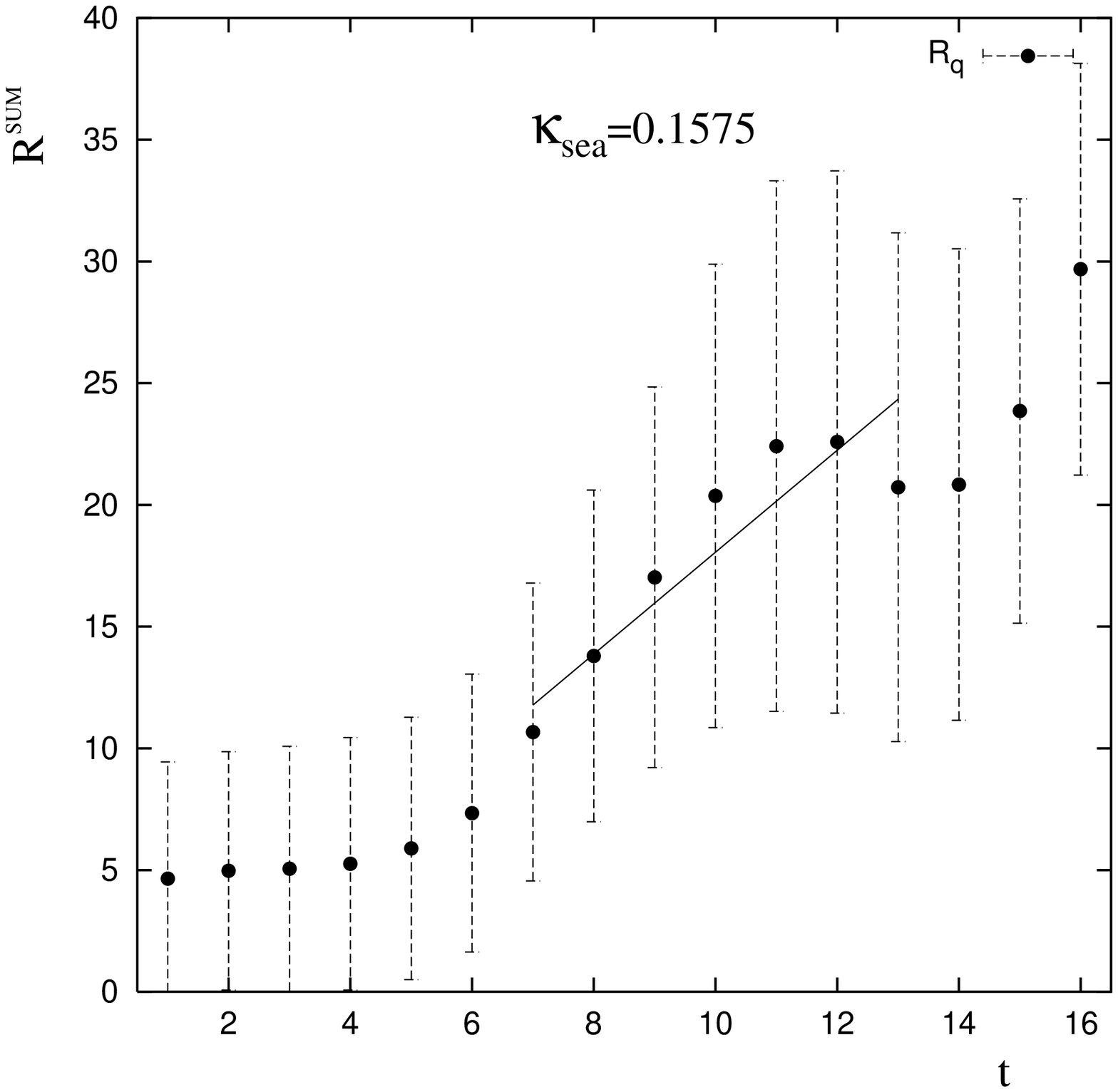}}\\
}
\caption{\label{fig_raw_disc_sum} {\it Summation method: The raw data $R_q$ for
the  disconnected amplitudes $D_q$ at our
sea quark masses. The fits (range and value) are indicated by solid lines.}}
\end{center}
\end{figure}
\begin{figure}[htb]
\begin{center}
\vskip -3.0cm
\noindent\parbox{15.0cm}{
\parbox{7.0cm}{\epsfxsize=7.0cm\epsfbox{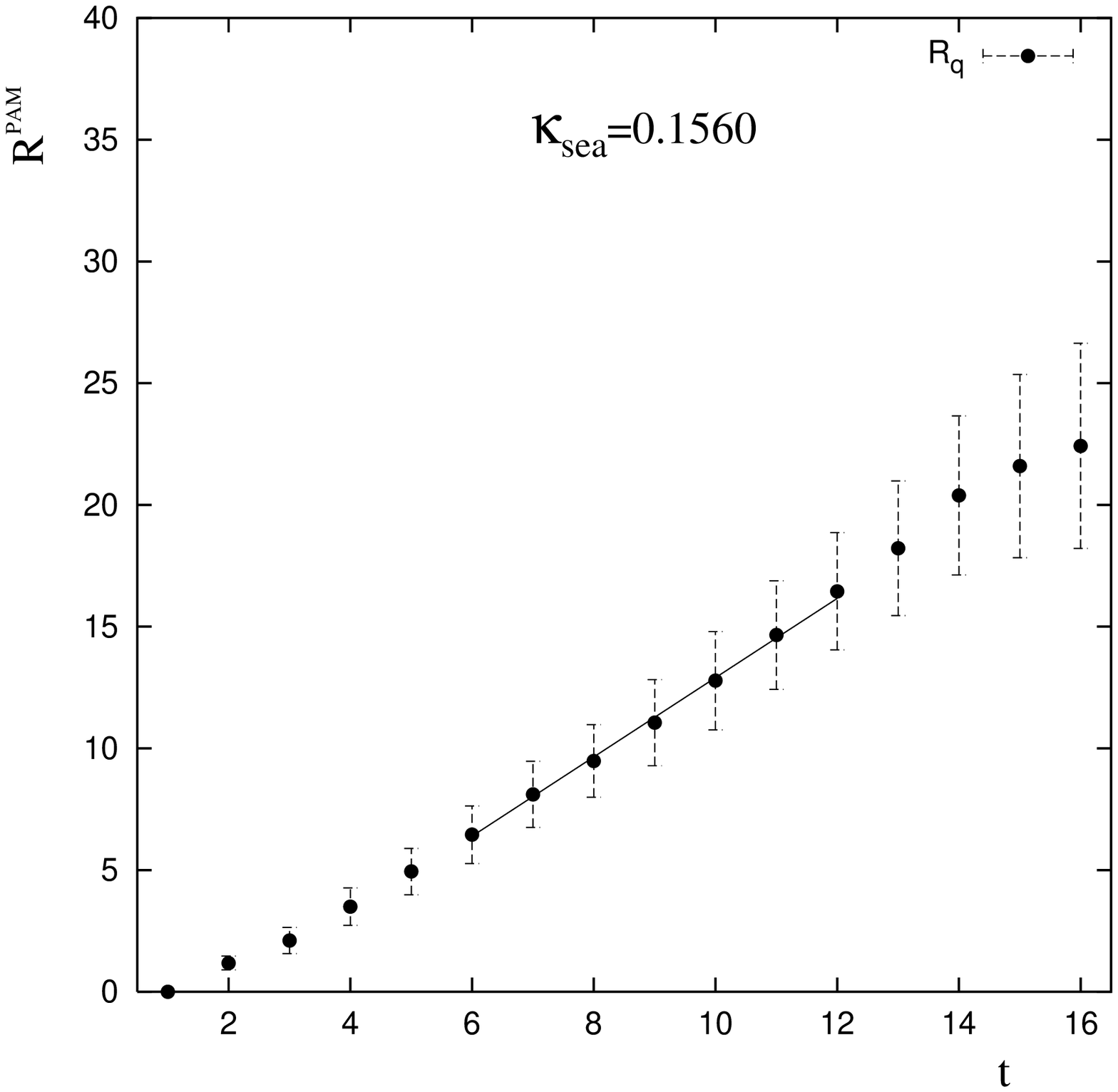}}
\parbox{7.0cm}{\epsfxsize=7.0cm\epsfbox{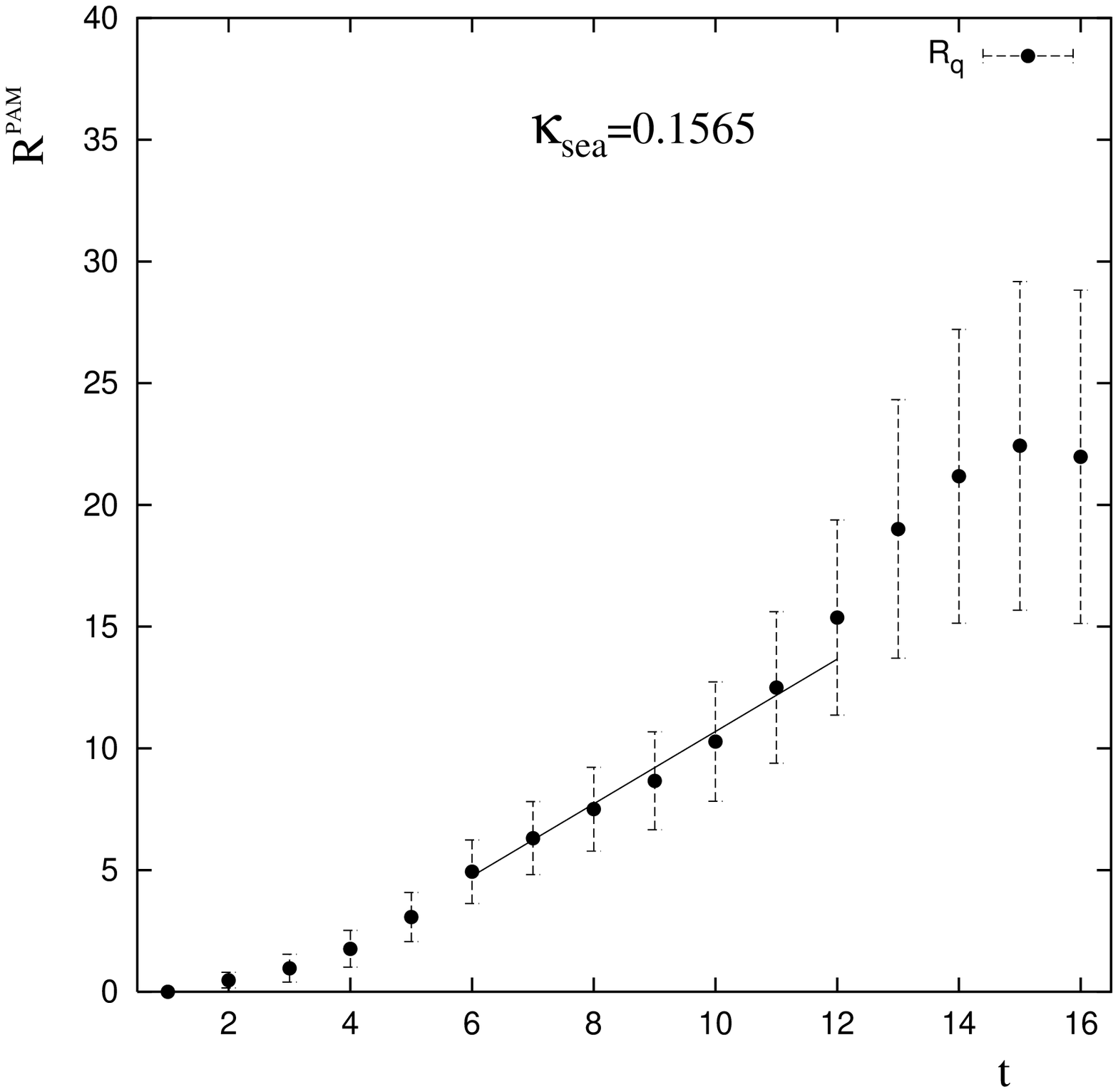}}
\\ \linebreak
\vskip -4.0cm
\parbox{7.0cm}{\epsfxsize=7.0cm\epsfbox{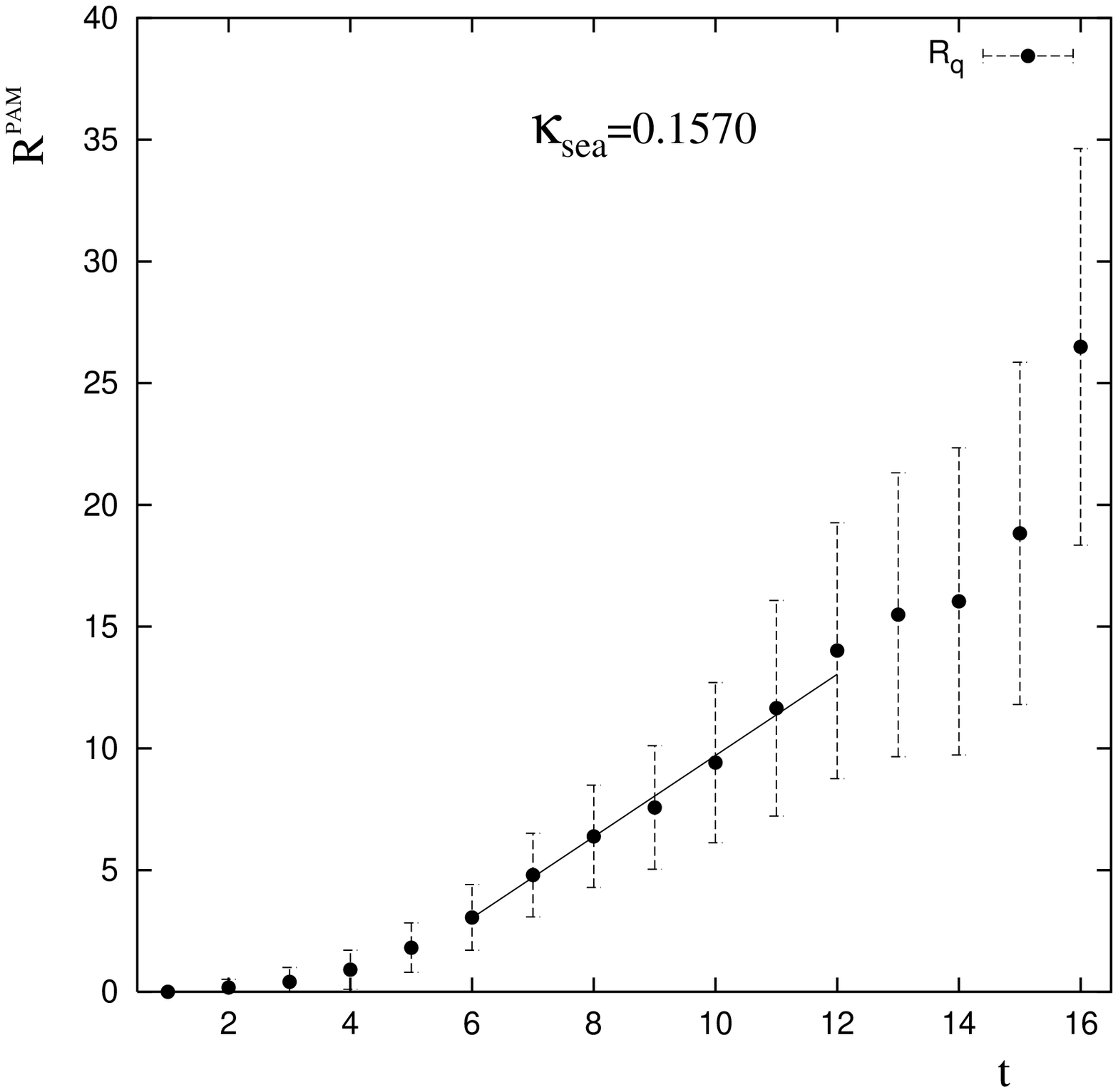}} 
\parbox{7.0cm}{\epsfxsize=7.0cm\epsfbox{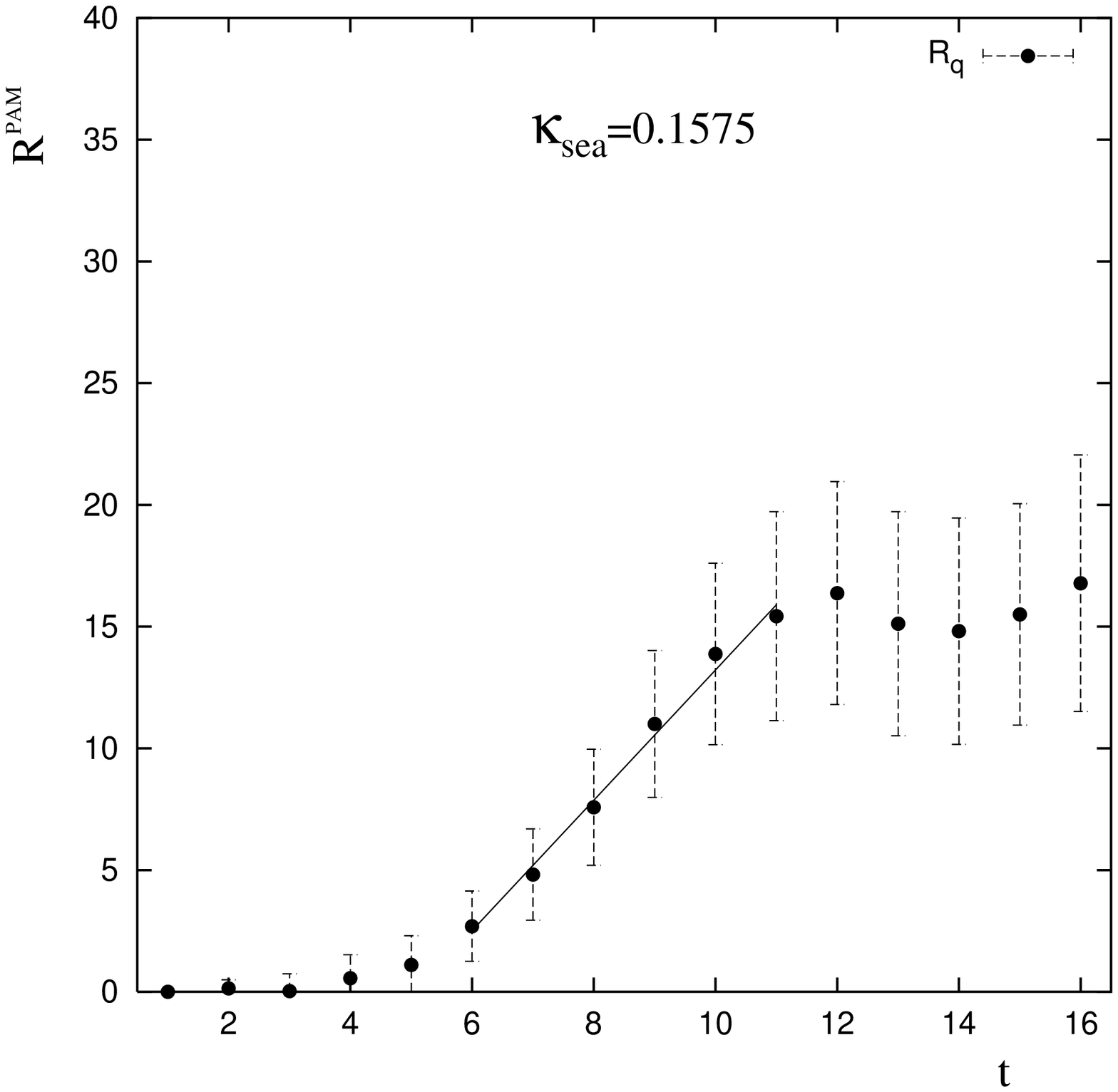}}\\
}
\caption{\label{fig_raw_disc_PAM} {\it PAM with
$\Delta {t_0}=\Delta t=1$: The raw data $R_q$ for
the  disconnected amplitudes $D_q$ at our
sea quark masses. The fits (range and value) are indicated by solid lines.}}
\end{center}
\end{figure}

The signals of the disconnected contributions are shown in 
fig. \ref{fig_raw_disc_sum} for the summation method and in 
fig. \ref{fig_raw_disc_PAM}   for PAM.

It turns out that the data analysed with the summation technique is
quite noisy and the signal tends to vanish at the smallest quark
mass. In contrast, PAM produces significantly improved signal to
noise ratios over our entire range of quark masses.

The outcome of the fits according to eqs.\ref{eq_sum_meth_asympt} and
\ref{eq_mplateau_meth_asympt} is compiled in 
tab.\ref{tab_raw_disconnected}.  The PAM data are clearly superior with
respect to statistical errors. It is gratifying to find the results of
both methods to be compatible within (large) statistical
uncertainties.

\begin{figure}[htb]
\begin{center}
\vskip -3.0cm
\noindent\parbox{15.0cm}{
\parbox{7.0cm}{\epsfxsize=7.0cm\epsfbox{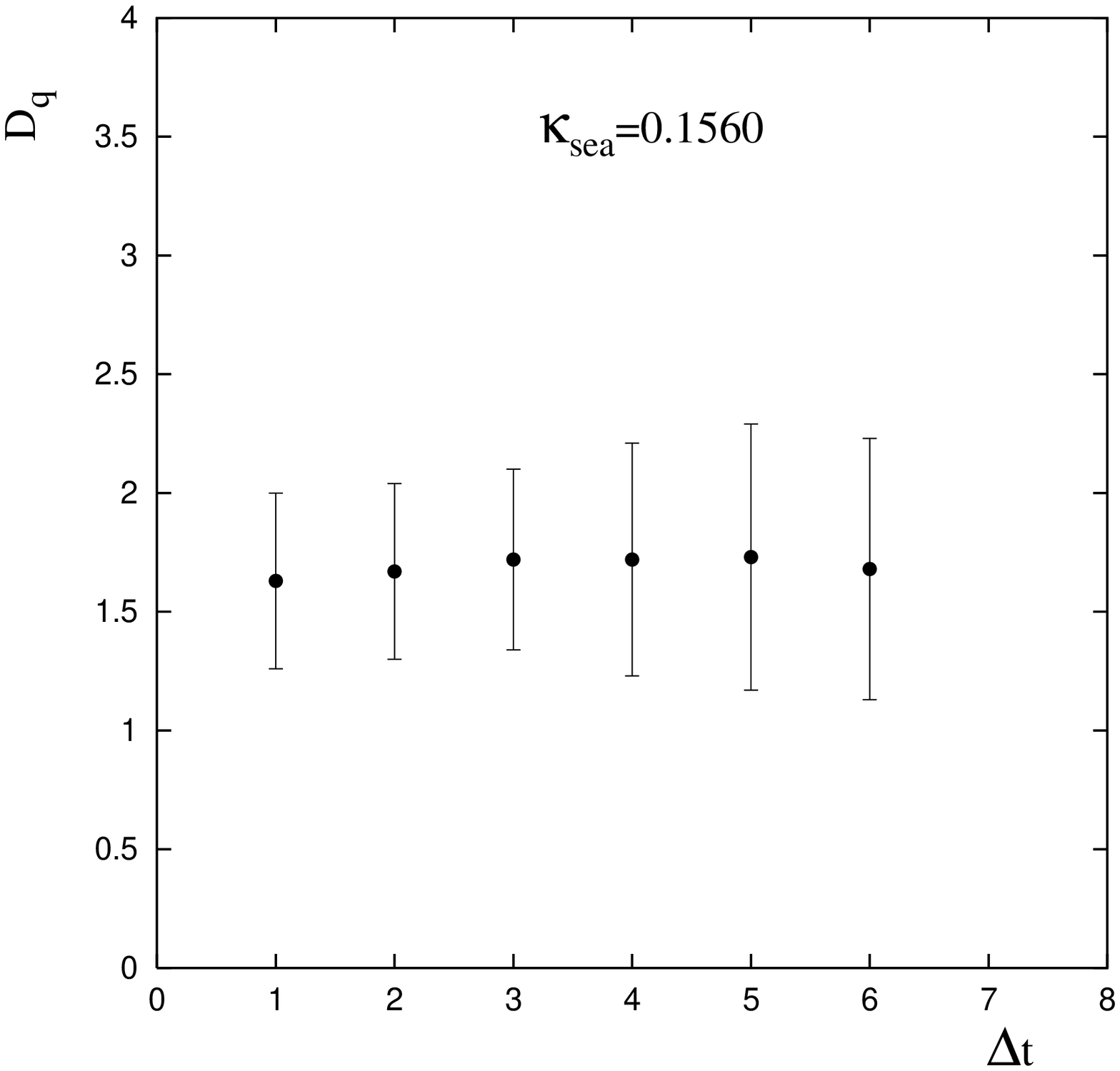}}
\parbox{7.0cm}{\epsfxsize=7.0cm\epsfbox{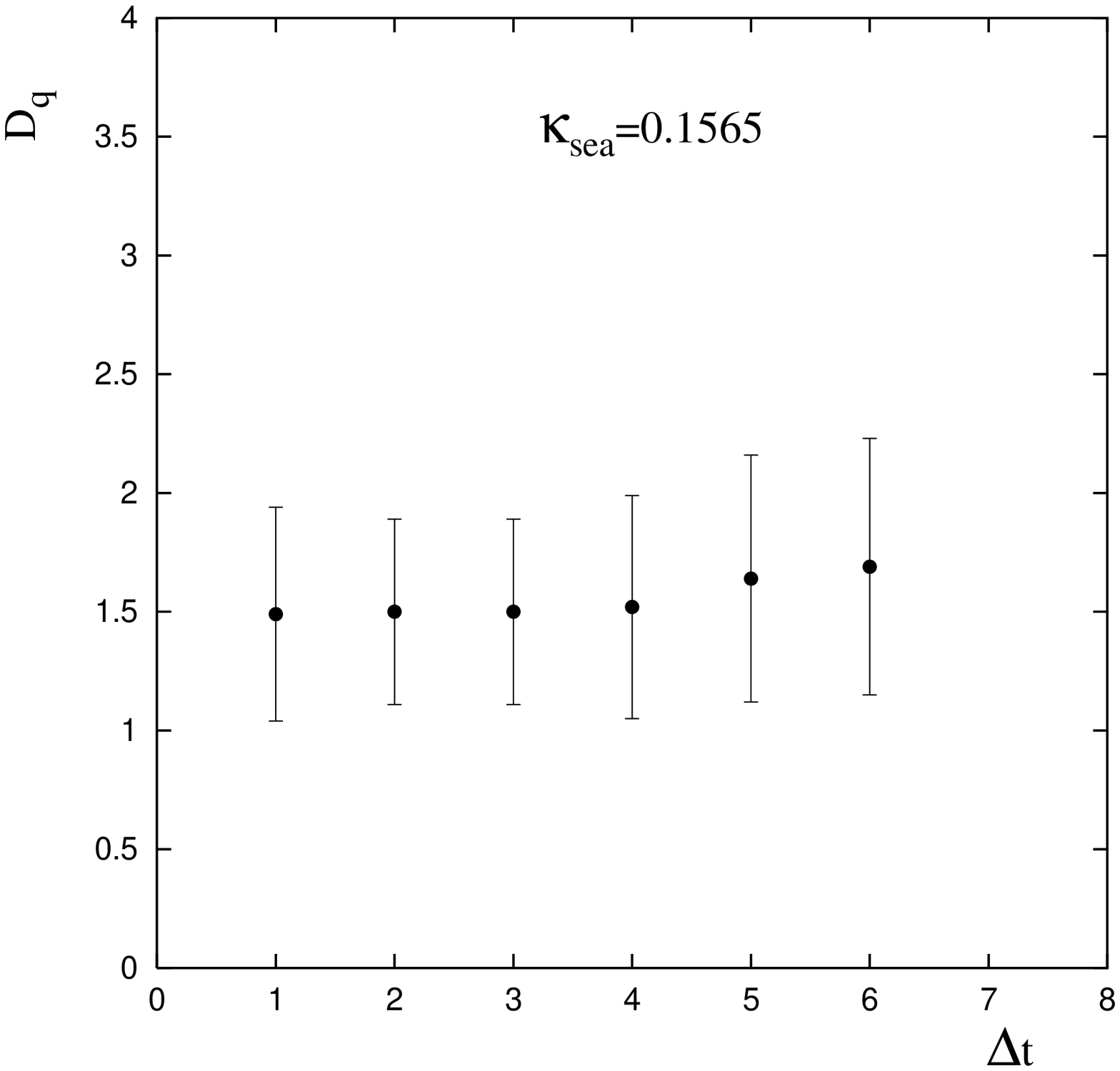}}
\\ \linebreak
\vskip -4.0cm
\parbox{7.0cm}{\epsfxsize=7.0cm\epsfbox{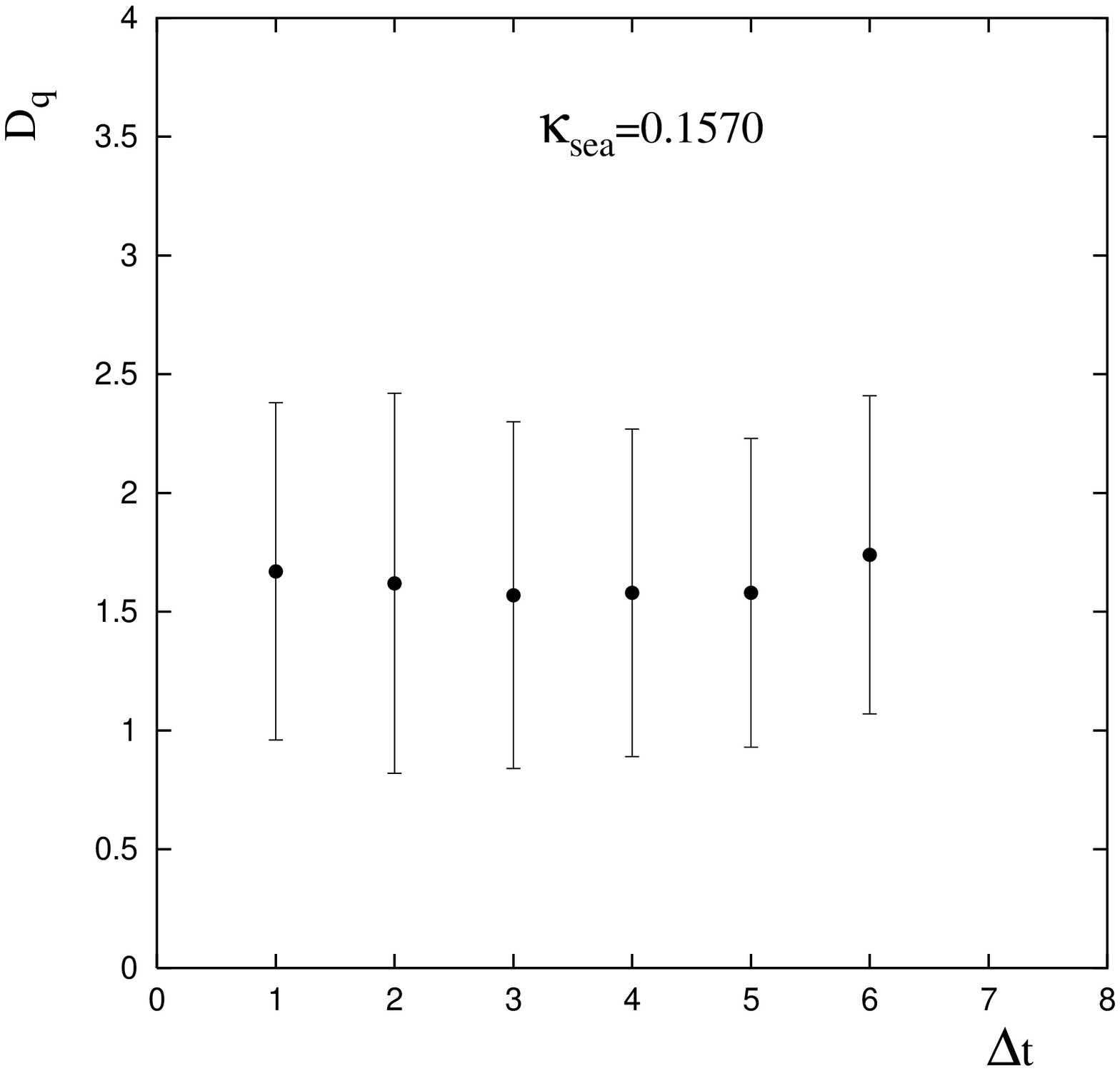}} 
\parbox{7.0cm}{\epsfxsize=7.0cm\epsfbox{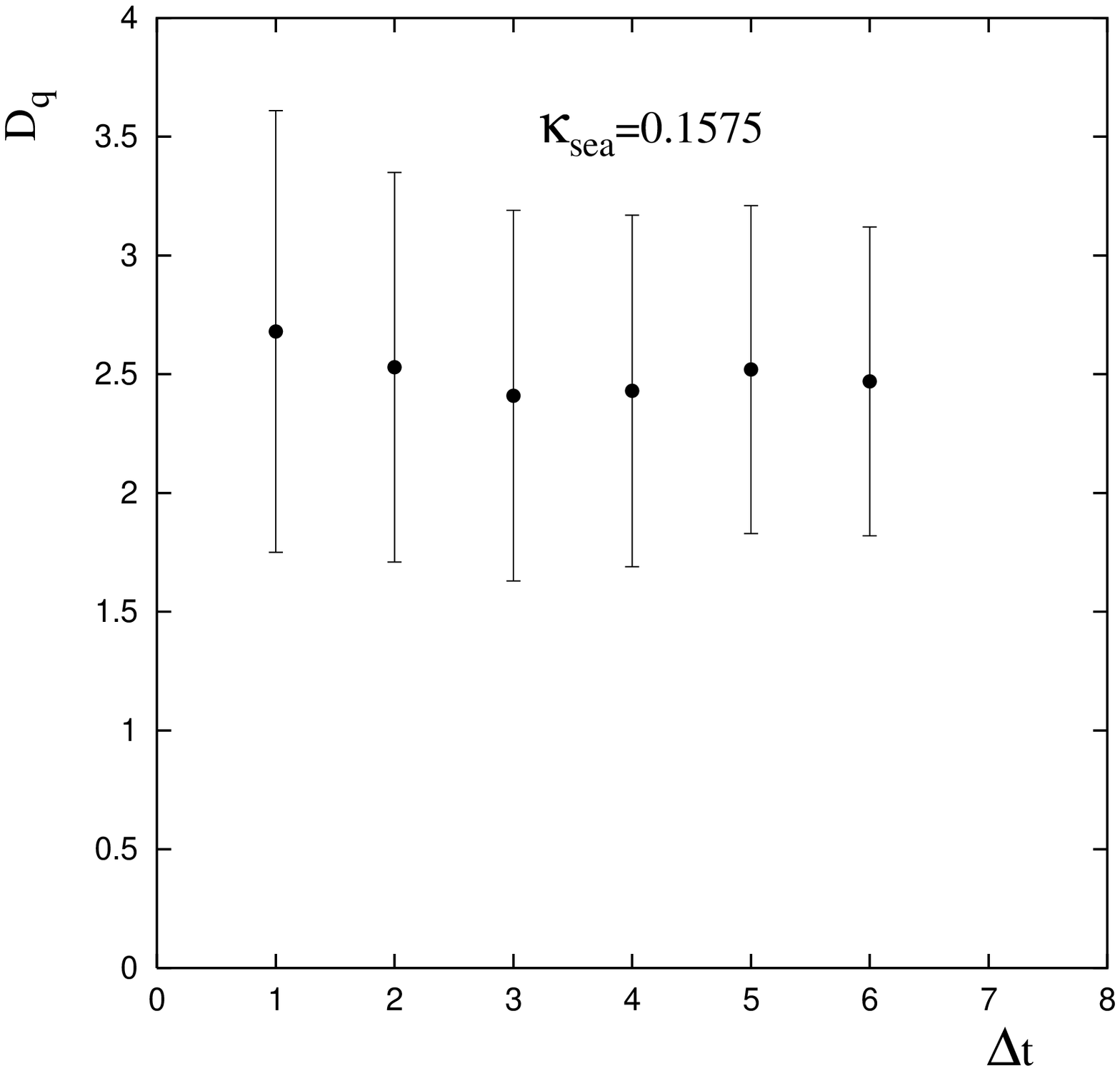}}\\
}
\caption{\label{fig_pam_dt} {\it PAM: The dependence of disconnected
amplitudes $D_q$ on $\Delta t$.
}}
\end{center}
\end{figure}

In order to assess the systematics of PAM we have evaluated
$R^{PAM}$ for several values of $\Delta t_0 = \Delta t$.
As can be seen from fig. \ref{fig_pam_dt}, $D_q$ shows no dependence on 
$\Delta t$, within our our statistical precision (200 gauge
configurations). Throughout this paper we
have therefore used $\Delta t_0 = \Delta t = 1$.

In tab.\ref{tab_raw_disconnected} we have also included the raw data
for the strange loop matrix element $\langle P|\bar{s}s|P\rangle$,
which will be used for the determination of the $y$ parameter.  To
obtain this data we held  the loop quark fixed at the
strange quark mass (c.f. \cite{sesam_light_spectrum}), and kept 
valence and sea quark masses degenerate under  chiral extrapolation.
\begin{table}[ht]
\begin{center}
\begin{tabular}{|c|c|c|c|}
\hline 
           &          &            &            \\
Method     & $\kappa$ & $D_q$      & $D_s$     \\
           &         &             &            \\ \hline
SUM        & 0.1560  & 0.86(54)    & 0.79(54)    \\
           & 0.1565  & 1.76(76)    & 1.74(74)     \\
           & 0.1570  & 3.22(1.68)  & 3.19(1.61)   \\
           & 0.1575  & 2.09(1.64)  & 1.85(1.44)   \\ \hline
           & 0.1560  & 1.63(37)    & 1.52(35)   \\
PAM      & 0.1565  & 1.49(45)    & 1.52(44)   \\
           & 0.1570  & 1.67(71)    & 1.64(68)   \\
           & 0.1575  & 2.68(93)    & 2.49(86)   \\ \hline
\end{tabular}
\caption{\label{tab_raw_disconnected}{\it Lattice results
(unrenormalised) of the disconnected amplitudes 
$D_q = \langle P(\kappa_{sea})
|\bar{q}q(\kappa_{sea})|P(\kappa_{sea})\rangle_{dis}$ and
$D_s = \langle P(\kappa_{sea})|\bar{q}q(\kappa_{str})|
P(\kappa_{sea})\rangle_{dis}$. $\kappa_{str}=0.15608$. }}
\end{center}
\end{table}
 
\section{Physics Results}

\subsection{Pion-Nucleon $\sigma$-Term}
  
Our procedure to obtain the physical value of $\sigma_{\pi N}$ is to
extrapolate the results of tabs.\ref{tab_raw_connected},
\ref{tab_raw_disconnected} with respect to the
quark mass to $m_{ud}$, and to multiply subsequently with the
lattice cutoff $a^{-1}$ and $m_{ud}$. Note that no renormalization
is necessary, as $\sigma_{\pi N}$ is a renormalization group invariant
quantity.
\begin{figure}[htb]
\begin{center}
{\epsfxsize=10.0cm\epsfbox{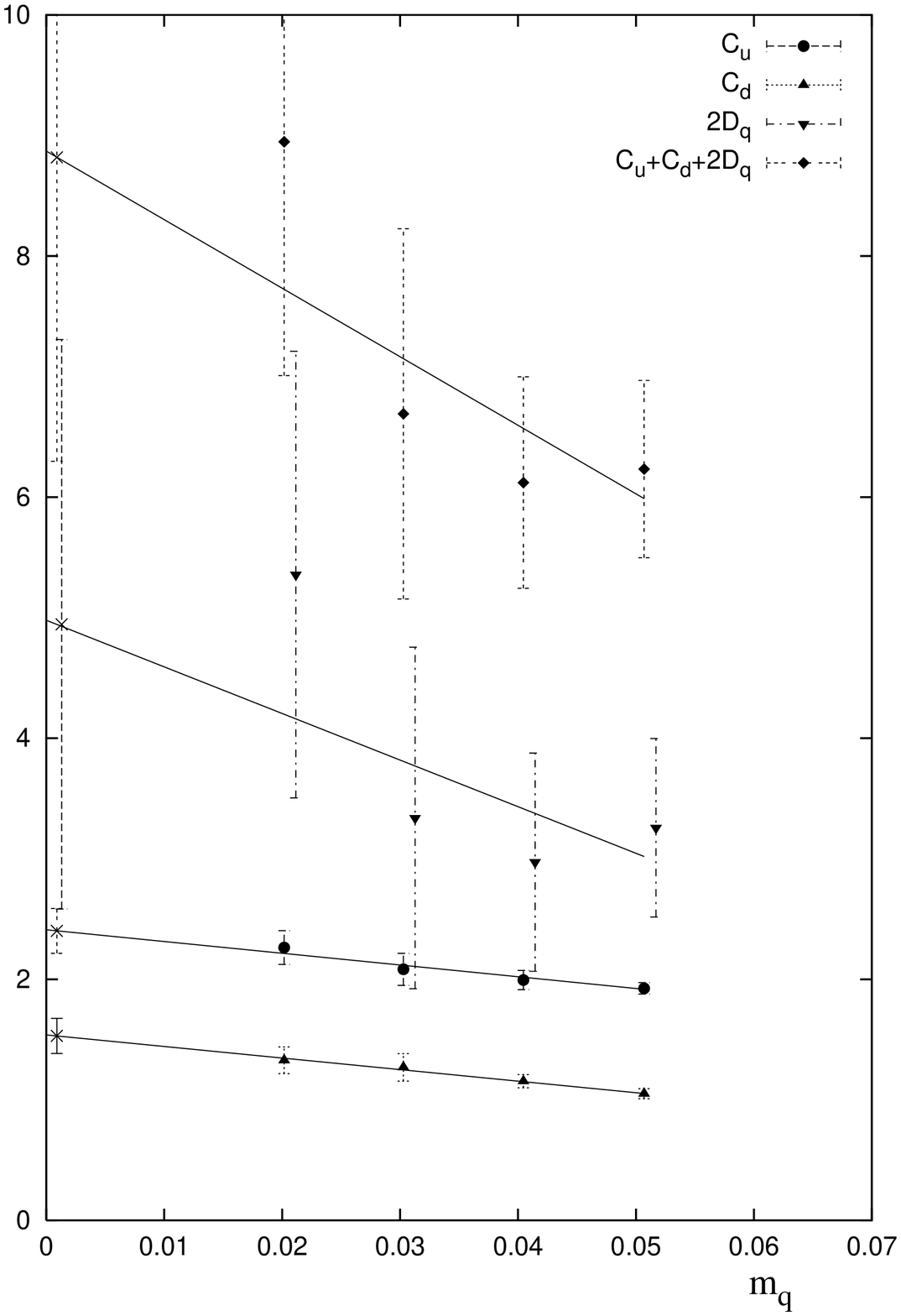}}
\caption{\label{fig_extrap_ins} {\it Linear extrapolations of connected and
disconnected amplitudes to the light quark mass. Bursts indicate the
results of the extrapolation.}}
\end{center}
\end{figure}

In fig.\ref{fig_extrap_ins} we display the extrapolations of the connected and
disconnected amplitudes. Since the statistical quality of the
disconnected contribution does not allow to resolve for higher orders
in $m_q$, we have decided to use consistently a linear ansatz for
all contributions. We emphasize that this is equivalent to a
quadratic ansatz in $M_N(m_q)$, since
the Feynman-Hellmann theorem yields 
$\partial M_N/\partial m_q(m_{ud}) = \langle P|\bar{u}u+\bar{d}d|P\rangle$.

The results of the extrapolations are collected 
in tab.\ref{tab_extrap_ins}.
\begin{table}[ht]
\caption{\label{tab_extrap_ins}{\it Lattice results for connected
and disconnected amplitudes at $\kappa_{light}=0.158462$. We have used
the data from the plateau method PAM for the connected and 
disconnected amplitudes respectively.}}
\begin{center}
\begin{tabular}{|c|c|c|c|}
\hline
$C_u$ & $C_d$ &  $2D_q$  & $C_u+C_d+2D_q$ \\
\hline
2.40(19) & 1.53(15) & 4.95(2.36) & 8.82(2.52)  \\
\hline
\end{tabular}
\end{center}
\end{table}
Note that
the ratio of disconnected to connected contributions at the light
quark is mass

\begin{equation}
R_{d/c} = \frac{\langle P|\bar{u}u + \bar{d}d|P \rangle_{disc}}
{\langle P|\bar{u}u + \bar{d}d|P \rangle_{con}} = \frac{2D_q}{C_u+C_d}
= 1.26(57)\;.
\label{eq_ratio_disc_con}
 \end{equation}
 
As a cross check we compare the
total value $C_u+C_d+2D_q=8.82(2.52)$ with our previous 
result from the quadratic
extrapolation of the nucleon mass\cite{sesam_light_spectrum}
\begin{equation}
\frac{\partial M_N}{\partial m_q}(m_{ud}) = 11.7(4.9)\;,
\end{equation} 
and find agreement. Note that the statistical uncertainty of the total
amplitude obtained with the direct (ratio) method is smaller than the
one from the derivative method (29$\%$ compared to 42$\%$). 

Finally, we determine the pion-nucleon $\sigma$-term in physical units.
With\footnote{We use the standard definition
of the 
quark mass $m_q = \frac{1}{2}(\frac{1}{\kappa} -
\frac{1}{\kappa_c})$.}
 $m_{ud} = 0.000901(54)$ (in lattice units) 
and $a_{\rho}^{-1}=2.30$GeV \cite{sesam_light_spectrum}
one obtains
\begin{equation}
\sigma_{\pi N} = a_{\rho}^{-1} m_{ud} \langle P|\bar{u}u +
\bar{d}d|P\rangle = 18(5)\mbox{MeV}\;.
\label{eq_sigma_result}
\end{equation} 
The statistical error has been determined with a full jackknife
analysis of the product, including the jackknife distributions of
$m_{ud}$ and $a_{\rho}^{-1}$.

Clearly, this is a rather small value, compared to the 
estimate from experiment. The latter is quoted
to be $\sigma_{\pi N}\simeq 45$MeV,
which is in rough agreement with quenched lattice
estimates. 

We emphasize that the origin for
this apparent dramatic unquenching effect on $\sigma_{\pi N}$ 
appears to be the substantial drop 
observed in the light quark mass $m_{ud}$: Our result
\cite{sesam_quarkmass}, $m_{ud}=2.7(2)$MeV, which has been obtained
using tadpole improved renormalization\cite{lepage_tadpole} 
\begin{equation}
Z_S = \frac{1}{2\kappa}\left( 1 - \frac{3 \kappa}{4 \kappa_c}\right)
\left[ 1 - 0.0098 \alpha_{\bar{MS}}(\frac{1}{a})\right]\;,
\label{eq_ZS} 
\end{equation}  
where $\alpha_{\bar{MS}}(\frac{1}{a})= 0.215$, 
is lower than the 
corresponding quenched estimate at equal cutoff ($a^{-1}\simeq 2.3$GeV),
$m_{ud} \simeq 5$MeV\cite{cp_pacs_burkhalter,cp_pacs_yoshie},
by roughly a factor of two.

Gupta has reasoned in the context of his discussion on light
quark masses\cite{gupta_lat97} that the validity of the one loop
formula, eq.\ref{eq_ZS}, is questionable in full QCD, insofar large
non perturbative contributions due to the presence of sea quarks could
arise and readily change $Z_m = Z_S^{-1}$ by a factor of two, thus
compensating the above factor 1/2 in $m_{ud}$.  However, this doubt
does not provide a source for a possible  underestimate to
$\sigma_{\pi N}$, since we are dealing here with a renormalization
group invariant quantity.

On the other hand, one might question the standard definition and
switch to  the lattice quark mass as defined by use of the axial vector
Ward identity\cite{gupta_lat97,rompisa,cp_pacs_yoshie}.
But from the final analysis of the CP-PACS group on quenched QCD reported
recently \cite{cp_pacs_yoshie} one would gather a {\it decrease} of $m_{ud}$
and $\sigma_{\pi N}$ by $20 - 50\%$!
One should remember, though, that both definitions of $m_{ud}$ differ at our
value of $a$ equally from 
their common continuum limit, and it is thus not a priori 
obvious which one of them is more suitable. 

Obviously the only way to decide on this issue is to perform a scaling
analysis of $m_{ud}$, i.e.  to repeat its calculation at several lattice
cutoffs $a^{-1}$, and then to extrapolate to the continuum
$a\rightarrow 0$. We mention that the CP-PACS collaboration has
launched  such a scaling study in full QCD, based on  three different
values of the cutoff\cite{cp_pacs_burkhalter}.  Their very preliminary
analysis strongly hints at  a {\it small value} of the light quark mass
in the continuum limit, which incidentally is close to the estimate
underlying our present work\cite{sesam_quarkmass}.
  
\subsection{$y$ Parameter}

The determination of the $y$ parameter, defined in eq.\ref{eq_y_def},
requires the calculation of the (purely disconnected) matrix element
$\langle N|\bar{s}s|N \rangle$. In principle, a fully consistent QCD
lattice analysis of this quantity can be done only in the setting of a
realistic $n_f \ge 3$ simulation, in which strange and light quarks
enter the dynamics through the fermion determinant.

Given our $n_f=2$ setup, we can only resort to a semi-quenched
analysis of the strange sector. This is done by identifying the masses
of the sea quarks with those of the light valence quarks of the
nucleon. The mass of the strange quark, which has no counterpart in
the sea, is contained to match the strange hadron spectrum in
semi-quenched analysis\cite{sesam_light_spectrum}. This procedure
accounts at least for the influence of light sea quarks on the above
matrix element and is certainly more adequate than previous quenched
($n_f=0$) calculations.

The lattice raw results $D_s=\langle
P(\kappa_{sea})|\left[\bar{s}s\right](\kappa_s)|P(\kappa_{sea})\rangle$
are listed in tab.\ref{tab_raw_disconnected}. The values differ only
slightly from the raw data $D_q$, indicating that an approximate
$SU(3)$ flavour invariance of the disconnected contributions is still
realised in our $n_f=2$ simulation.

We extrapolate $D_s$ as a function of $\kappa_{sea}$ linearly to the
light quark mass.  To obtain the $y$ parameter we renormalise our
result
\begin{equation}
2\langle P|\bar{s}s|P \rangle = 4.94(2.14) 
\end{equation}
by $Z_s$, eq.\ref{eq_ZS}. This yields
\begin{equation}
y = \frac{2 Z_S(\kappa_s)\langle P|\bar{s}s|P\rangle}
{Z_s(\kappa_l)\langle P|\bar{u}u + \bar{d}d|P\rangle} = 0.59(13) \;.
\label{eq_y_result}
\end{equation} 
 

It has been suggested by Laga\"{e} and Liu \cite{liu_z_factors} that
the lattice quark mass dependent part of the renormalization factor
would be quite different for connected and disconnected amplitudes.
According to their reasoning, based on first order lattice
perturbation theory, they would lower our value of $y$ by about
$30\%$.  We do however not agree with their arguments, at least in the
case of scalar quark loops, for the following reason: Since the
combination $m_q \langle P|\bar{q}q|P\rangle$ is a renormalization
group invariant quantity, both contributions to the amplitude,
connected {\it and} disconnected, will be renormalised equally, i.e.
by one and the same factor $Z_S = 1/Z_m$, where $Z_m$ renormalises the
quark mass. Choosing different factors, $Z_S^{con} \neq Z_S^{disc}$, as
they propose, would be in conflict with the renormalisation
group invariance.
  
\section{Discussion and Conclusions}

We have studied connected and disconnected contributions to the proton
scalar density amplitude in a $n_f=2$ full QCD simulation at fixed 
lattice cutoff and volume. 

It turns out that conventional ratio methods (summation and plateau)
yield excellent signals for the connected parts, but fail in the
determination of disconnected contributions.  The latter situation can
be improved substantially by use of PAM.
It is likely that stochastic estimator techniques
will perform even better on larger lattices, where we would
expect  self averaging
effects to reduce the errors from the present level, which is
30 - 40$\%$.

We find the ratio of disconnected to connected contributions, 
$R_{d/c}=1.26(57)$, at
$m_{ud}$ to be lower than the estimate from the quenched simulation
at $\beta=5.7$ of \cite{japan_nsigma}, where they quote $R_{d/c}=2.23(52)$.  
Albeit $\beta=5.7$ corresponds to rather strong coupling, this
difference in $R_{d/c}$ might indicate an unquenching effect.
In order to consolidate this tentative conclusion, one should (a) repeat
the analysis on a larger lattice, where self averaging will help to
reduce the statistical errors, and (b) carry out a quenched reference
simulation at equal lattice spacing and statistics. Work along this
line is in progress.


Due to the small value of $m_{ud}$ in full QCD, the
physical result for the pion-nucleon $\sigma$-term comes out rather
low. It remains to be seen whether this is a finite cutoff effect.
In this context it is very interesting that the preliminary
results of the CP-PACS collaboration\cite{cp_pacs_burkhalter}, obtained
in full QCD with an improved
action,  point at an increase of $m_{ud}$ by about 30$\%$ when
changing from the standard to the Ward identity definition, at our
value of $a$.   
It is obvious that a scaling analysis of the quark mass in full QCD is
of utmost importance. 

So far the $y$ parameter has only been determined with a
semi-quenched treatment of light quarks. The result, $y=0.59(13)$,
 is far away from the phenomenological expectation, $y \simeq 0.2 - 0.4$.
A full QCD calculation of $y$
with $n_f \geq 3$ dynamical flavours is therefore highly desirable.
 
\paragraph{Acknowledgments}
We thank G.~Ritzenh\"ofer for his contributions during the initial
stage of this analysis. This work has been supported by the DFG grants Schi
257/1-4, 257/3-2, 257/3-3 and by the DFG Graduiertenkolleg
``Feldtheoretische Methoden in der Statistischen und
Elementarteilchenphysik''. The connected contributions have been
computed on the CRAY T3E systems of ZAM at FJZ. The disconnected parts
were calculated on the APE100 computers at IfH Zeuthen and on the
Quadrics machine provided by the DFG to the Schwerpunkt "`Dynamische
Fermionen"', operated by the University of Bielefeld. We thank the
staffs of these institutions for their kind support.

\end{document}